\documentclass[aps,twocolumn,longbibliography,superscriptaddress]{revtex4-2}
\usepackage{amsmath,amssymb,mathrsfs}
\usepackage{natbib}
\usepackage{subfigure}
\usepackage{tabularx}
\usepackage{epsfig}
\usepackage{longtable}
\usepackage{amsfonts}
\usepackage{rotating}
\usepackage{subfigure}
\usepackage{amsmath}
\usepackage{comment}
\usepackage{bbold} 
\usepackage{multirow}
\usepackage{hhline}
\usepackage{color}

\def\be{\begin{equation}}
\def\ee{\end{equation}}
\def\bea{\begin{eqnarray}}
\def\eea{\end{eqnarray}}
\def\nn{\nonumber}

\def\up{\uparrow}
\def\down{\downarrow}

\newcommand{\hc}[0]{\text{h.c.}}
\newcommand{\bra}[1]{ \left\langle #1 \right\vert }
\newcommand{\ket}[1]{ \left\vert #1 \right\rangle }

\usepackage[unicode=true,bookmarks=true,bookmarksnumbered=false,bookmarksopen=false,breaklinks=false,pdfborder={0 0 1},backref=false,colorlinks=true]{hyperref}

\hypersetup{linkcolor=magenta,urlcolor=blue,citecolor=blue,pdfstartview={FitH},hyperfootnotes=false,unicode=true}

\begin{document}

\title{Universal Quantum Optimization with Cold Atoms in an Optical Cavity} 
\author{Meng Ye$^\S$}
\affiliation{State Key Laboratory of Surface Physics, Key Laboratory of Micro and Nano Photonic Structures (MOE), and Department of Physics, Fudan University, Shanghai 200433, China}
\affiliation{Shanghai Qi Zhi Institute, AI Tower, Xuhui District, Shanghai 200232, China} 
\author{Ye Tian$^\S$} 
\affiliation{Department of Physics and State Key Laboratory of Low Dimensional Quantum Physics, Tsinghua University, Beijing, 100084, China}
\affiliation{Frontier Science Center for Quantum Information, Beijing, 100084, China} 
\affiliation{Collaborative Innovation Center of Quantum Matter, Beijing, 100084, China} 
\author{Jian Lin}
\affiliation{State Key Laboratory of Surface Physics, Key Laboratory of Micro and Nano Photonic Structures (MOE), and Department of Physics, Fudan University, Shanghai 200433, China}  
\author{Yuchen Luo}
\affiliation{State Key Laboratory of Surface Physics, Key Laboratory of Micro and Nano Photonic Structures (MOE), and Department of Physics, Fudan University, Shanghai 200433, China}
\affiliation{Shanghai Qi Zhi Institute, AI Tower, Xuhui District, Shanghai 200232, China} 
\author{Jiaqi You}
\affiliation{Department of Physics and State Key Laboratory of Low Dimensional Quantum Physics, Tsinghua University, Beijing, 100084, China}
\affiliation{Frontier Science Center for Quantum Information, Beijing, 100084, China} 
\affiliation{Collaborative Innovation Center of Quantum Matter, Beijing, 100084, China} 
\author{Jiazhong Hu} 
\affiliation{Department of Physics and State Key Laboratory of Low Dimensional Quantum Physics, Tsinghua University, Beijing, 100084, China}
\affiliation{Frontier Science Center for Quantum Information, Beijing, 100084, China} 
\affiliation{Collaborative Innovation Center of Quantum Matter, Beijing, 100084, China} 
\author{Wenjun Zhang}
\email{zhangwenjun@ultracold.group} 
\affiliation{Department of Physics and State Key Laboratory of Low Dimensional Quantum Physics, Tsinghua University, Beijing, 100084, China}
\affiliation{Frontier Science Center for Quantum Information, Beijing, 100084, China} 
\affiliation{Collaborative Innovation Center of Quantum Matter, Beijing, 100084, China} 
\author{Wenlan Chen} 
\email{cwlaser@ultracold.cn} 
\affiliation{Department of Physics and State Key Laboratory of Low Dimensional Quantum Physics, Tsinghua University, Beijing, 100084, China}
\affiliation{Frontier Science Center for Quantum Information, Beijing, 100084, China} 
\affiliation{Collaborative Innovation Center of Quantum Matter, Beijing, 100084, China} 
\author{Xiaopeng Li} 
\email{xiaopeng\underline{ }li@fudan.edu.cn} 
\affiliation{State Key Laboratory of Surface Physics, Key Laboratory of Micro and Nano Photonic Structures (MOE), and Department of Physics, Fudan University, Shanghai 200433, China} 
\affiliation{Shanghai Qi Zhi Institute, AI Tower, Xuhui District, Shanghai 200232, China} 
\affiliation{Institute of Nanoelectronics and Quantum Computing, Fudan University, Shanghai 200433, China}
\affiliation{Shanghai Artificial Intelligience Laboratory, Shanghai 200232, China} 
\affiliation{Shanghai Research Center for Quantum Sciences, Shanghai 201315, China}

\begin{abstract}
Cold atoms in an optical cavity have been widely used for quantum simulations of many-body physics, where the quantum control capability has  been advancing rapidly in recent years. Here, we show the atom cavity system is universal for quantum optimization with arbitrary connectivity. We consider a single-mode cavity and develop a Raman coupling scheme by which  the engineered quantum Hamiltonian for atoms directly encodes number partition problems (NPPs). The programmability is introduced by placing the atoms at  different positions in the cavity with optical tweezers. The  NPP solution is encoded in the ground state of atomic qubits coupled through a photonic cavity mode, that can be reached by adiabatic quantum computing (AQC).  We construct an explicit mapping for the 3-SAT and vertex cover problems to be efficiently encoded by the cavity system, which costs  linear overhead in the number of atomic qubits. The atom cavity encoding is further extended to  quadratic unconstrained binary optimization  (QUBO) problems.  The  encoding protocol is optimal in the cost of atom number scaling with the number of binary degrees of freedom of the computation problem. Our theory implies the atom cavity system is a promising quantum optimization platform searching for practical quantum advantage. 
\end{abstract}

\maketitle

\paragraph*{Introduction.---}Quantum optimization aims at utilizing  quantum fluctuations to solve difficult binary optimization problems. The idea is to encode the computation solution into  the ground state of certain programmable quantum many-body Hamiltonian systems. Their ground states can be prepared using quantum adiabatic or variational principles, for example with AQC~\cite{2001_Farhi_Science} or quantum approximate optimization algorithms (QAOA)~\cite{2014_Farhi_QAOA}. 
It has wide applications including  protein folding~\cite{2012_Aspuru-Guzik_SciRep}, simulating molecular dynamics~\cite{2022_Gaidai_MD}, and modeling wireless communication networks~\cite{1990_Unit_disk}.
While quantum optimization may not solve NP-complete or NP-hard problems at polynomial costs, it is widely expected to exhibit significant quantum speedup over classical computing~\cite{2020_Zhou_PRX,2021_Hastings_Quantum}, and recent studies have shown the quantum dynamics are less vulnerable than classical searching algorithms to trapping at local minima, a standard obstacle for finding the optimal solution~\cite{2022_Lukin_Science,2023_Dwave_Nature}.  Quantum optimization protocols could {benefit from} even more drastic quantum speedup with machine learning based quantum algorithm configuration~\cite{lin2020quantum,lin2021hard,2022_Cirac_VQAA,2022_Hsieh_NMI}. 

A fascinating  route to implement quantum optimization has been provided by Rydberg atom arrays~\cite{2020_Browaeys_NatPhys}. This atomic system naturally encodes maximum independent set problems on unit disk graphs~\cite{1990_Unit_disk}. A quantum wiring approach has been developed to mediate arbitrary connectivity, which makes  the Rydberg atomic system  a generic QUBO solver despite its finite interaction range~\cite{2020_Qiu_PRXQ, 2023_Pichler_PRXQ}. 
Tremendous research efforts have been devoted to this system in recent years with remarkable progress accomplished~\cite{2021_Browaeys_Nature,2021_Lukin_Rydberg,2022_Ahn_NatPhys,2022_Saffman_Nature,2022_Sheng_PRL,2022_Lukin_Science}. The Rydberg system has become a prominent platform to achieve quantum advantage in practical applications~\cite{2020_Qiu_PRXQ,2022_Ahn_NatPhys,2022_Saffman_Nature,2022_Lukin_Science,2023_Pichler_PRXQ}. However, one key limitation of this system is its quantum coherence time 
for the encoding Rydberg qubits involve highly excited atomic states, whose quantum coherence is fundamentally affected by blackbody radiation~\cite{2010_Saffman_RMP} and electromagnetic noise~\cite{2017_Cappellaro_RMP}.

\begin{figure*}[htp]
\begin{center}
\includegraphics[width=.65\linewidth]{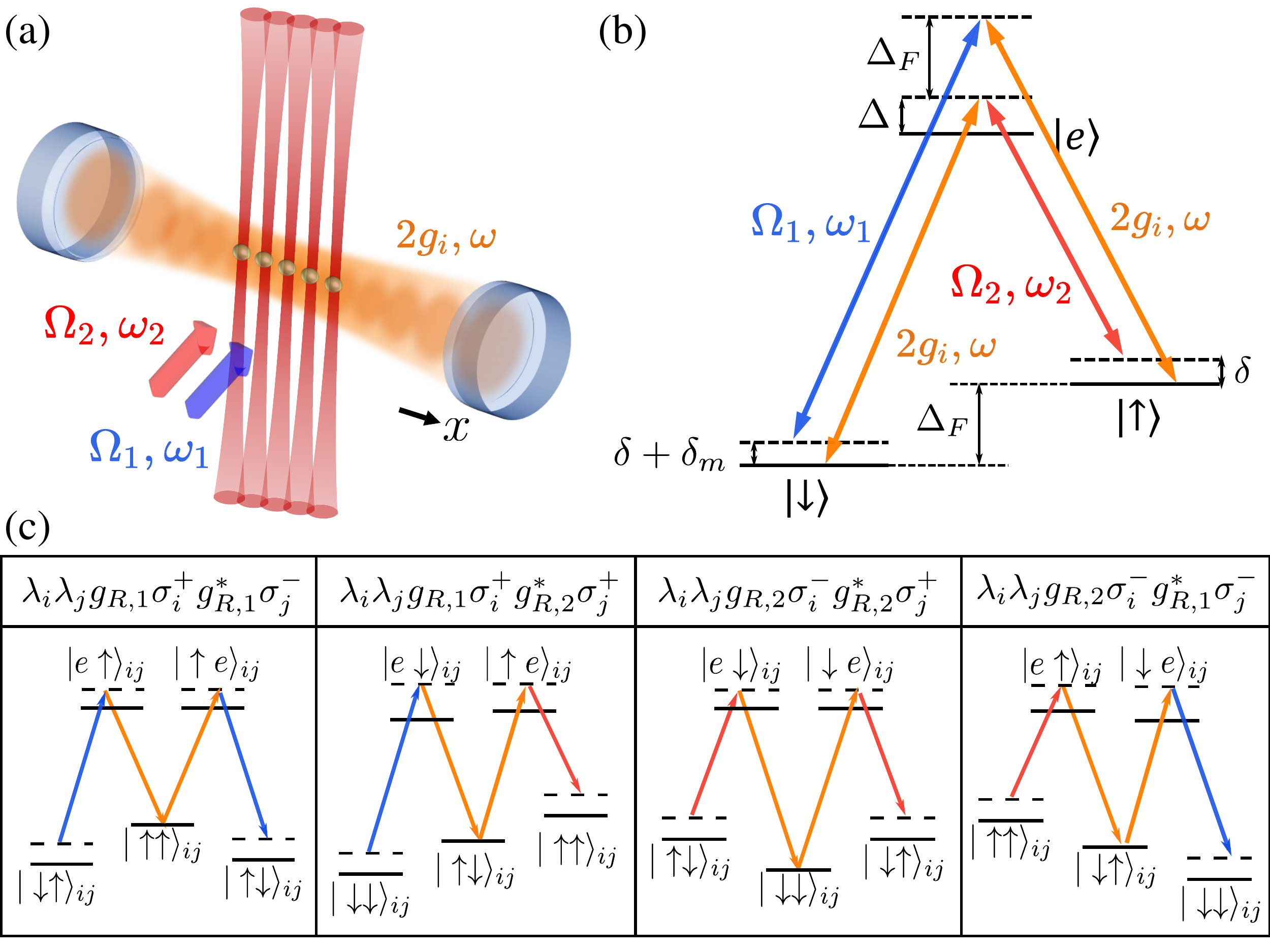}
\end{center}
\caption{\label{fig:experiment} Cavity QED setup for solving number partition problem. (a) Atoms are coupled to a high finesse optical cavity. The single-photon Rabi frequency $2g_i$ for the $i$-th atom is individually programmable by controlling the position of the atom with an optical tweezer. The atoms are illuminated by two additional beams, which are in a plane transverse to cavity axis, to generate the two Raman couplings. (b) The atomic level diagram. The qubit levels $\ket{\uparrow}$ and $\ket{ \downarrow}$ are coupled to $\ket{e}$ by the same cavity mode with detunings $\Delta+\Delta_F$ and $\Delta$, respectively. By applying two coupling beams $\Omega_1, \Omega_2$, and setting their detunings as $\Omega_1/\Omega_2=(\Delta+\Delta_F)/\Delta$, we construct two Raman processes that couple $\ket{\up} $ to $\ket{\down}$ with equal two-photon Rabi frequencies $g_{R}$. The detuning $\delta$ is chosen to be much larger than $g_R$ to 
suppress transitions generating multiple cavity photons. 
As we tune $\delta_m$ to 0, the four-photon processes become resonant. (c) Diagram of the resonant four-photon processes. These consist of two Raman processes with opposite detuning, and correspond to  $\hat\sigma^+_i \hat\sigma^+_j ,\hat\sigma^-_i \hat\sigma^-_j ,\hat\sigma^+_i \hat\sigma^-_j $, and $\hat\sigma^-_i \hat\sigma^+_j $. }
\end{figure*}

In this work, we consider a system of cold atoms in an optical cavity, whose experimental technology has been advancing rapidly in recent years~\cite{2013_Esslinger_RMP,2015_Rempe_RMP,2021_Wallraff_RMP,2019_Davis_PRL,2020_James_Nature,2021_Monika_Nature,2023_Tiancai_PRL}, 
and develop a novel quantum optimization scheme for generic QUBO problems with arbitrary connectivity using atomic ground states. 
The long-rang atomic  interactions mediated by cavity photons in this system naturally encode NPPs, having direct  applications in multi-processor scheduling of parallel computations and large scale truck delivery management~\cite{2006_Ukovich_NPPapplication}. 
Introducing a series of auxiliary squarefree integers, we show that 3-SAT and vertex cover  problems can be efficiently encoded by the atom cavity system, where the cost of atom number scales {\it linearly} with number of binary degrees of freedom of the computation problem, in contrast to the {\it quadratic} scaling in the corresponding  Rydberg encoding~\cite{2020_Qiu_PRXQ, 2023_Pichler_PRXQ}. 
Based on this scheme, we construct an encoding scheme where QUBO problems with {\it arbitrary}  connectivity are mapped to the atom cavity system. 
The overhead in the cost of atom number in representing QUBO problems  has a quadratic scaling in our scheme, which is already optimal. With our theory, the atom cavity system has a potential to become a universal quantum optimization platform to demonstrate practical quantum advantage.


\paragraph*{Solving NPP by cold atoms in an optical cavity.}
Given a set of $n$ positive integers, $ {\cal S} = \{p_1,p_2,...,p_n\}$, NPP is to  divide  the set into two subsets $A$ and $\overline{A}$ in order to minimize the imbalance 
$
I = \left|\sum_{i\in A} p_i - \sum_{j\in \overline{A}} p_j \right|. 
$
In order to map this computation problem to a quantum system, we rescale the integers in ${\cal S}$ by their maximum $p_{\rm max}$, and introduce $\lambda_j = p_j/p_{\rm max}$. 
The solution to NPP corresponds to the ground state of an Ising Hamiltonian, 
\be 
\textstyle \hat{H}_{\rm NPP} = 
\left(\sum_{i=1}^{n}  \lambda_i \hat\sigma_i^x\right)^2,
\label{eq:npphamiltonian}
\ee 
where $\hat\sigma_i^x$ is the Pauli-x matrix for the $i$-th qubit. The measurement of $\hat\sigma^i_x$ being positive (negative) encodes the $i$-th integer, $p_i$, given to $A$ ($\overline{A}$).  
The physical implementation requires the coupling between two qubits, say as labeled by $i$ and $j$, to be $\lambda_i \lambda_j$, which is nonlocal.

The required Hamiltonian $\hat H_{\rm NPP}$ with nonlocality has natural realization with atoms interacting with cavity photons. 
The key idea is to utilize a cavity-mediated four-photon process to realize the desired interactions in Eq.~\eqref{eq:npphamiltonian}. 
We consider $N$ three-level atoms in a high-finesse optical cavity (Fig.~\ref{fig:experiment}). 
The qubit is encoded by the two ground states $\ket{ \uparrow  }$ and $\ket{\downarrow} $ with an intrinsic energy splitting $\hbar\Delta_F$. Both ground states are coupled to the excited state $\ket{e}$ by the cavity photons $\hbar\omega$, with detunings of $\Delta+\Delta_F$ and $\Delta$, respectively.
The atoms are individually trapped by optical tweezers with programmable relative positions to the standing-wave cavity mode, leading to programmable atom-cavity coupling strengths. 
Denoting the single-photon Rabi frequency at anti-nodes as $2g_0$, the coupling strength of the $i$-th atom at position ${X}_i$ is given by $2g_i = 2g_0 \sin(QX_i)$, with $Q$ the cavity-mode wave-vector. By manipulating the atom positions, we reach a programmable coupling $g_i=\lambda_i g_0$.

In order to realize the Raman coupling between $\ket{\downarrow}$ and $\ket{\uparrow}$, we send in two side coupling beams to off-resonantly couple $\ket{ \downarrow}$ ($\ket{ \uparrow } $) to $|e\rangle$ with frequencies $\omega_1$ ($\omega_2$) and Rabi frequencies $\Omega_1$ ($\Omega_2$), as shown in Fig.~\ref{fig:experiment}(b).
The Hamiltonian $\hat H_{\rm exp}$ of such system is given by
\bea 
  \label{eq:experimental hamiltonian}
    &&\hat{H}_{\rm exp}/\hbar = \omega \hat{a}^\dagger \hat{a} - \sum_{i=1}^{N} \left\{ \omega_0 \ket{\down}_i\!\bra{\down} + (\omega_0-\Delta_F)\ket{\up}_i\!\bra{\up} \right\} \nn \\
    &&\enspace+ \sum_i \left\{ \Omega_1 \cos(\omega_1 t) \ket{e}_i\!\bra{\down} + \Omega_2 \cos(\omega_2 t) \ket{e}_i\!\bra{\up} +\hc \right\} \nn \\
    &&\enspace+ \sum_i \lambda_i g_0 \left\{ \hat{a}  \ket{e}_i\!\bra{\down} + \hat{a} \ket{e}_i\!\bra{\up} +\hc \right\}, 
\eea 
where $\hat a^\dagger$ ($\hat a$) is the creation (annihilation) operator of the cavity mode. Here 
$\Delta_F$ is the hyperfine splitting between the two ground states $\ket{\up} $ and $\ket{ \down} $, 
the energy zero point is set at the level of $\ket{e} $, $\omega$ is the cavity mode frequency, 
and $\omega_0$ is the frequency of the transition $\ket{\down} \rightarrow \ket{e} $. 
 The detunings are organized as
$\textstyle  \omega_1 = \omega_0 + \Delta + \Delta_F - \delta - \delta_m$, 
$\textstyle \omega_2 = \omega_0 + \Delta - \Delta_F - \delta$, 
and 
 $ \textstyle \omega  = \omega_0 + \Delta$.  
 The atom number $N$ is equal to the number of integers ($n$) to divide for NPP.

After rotating wave approximation and adiabatically eliminating the excited state (Supplementary Material), each side coupling beam 
{combined with} 
the cavity mode forms a detuned Raman coupling between the two ground states, described by the Hamiltonian, 
\bea 
    \hat{H}_{\rm exp}^{\prime}/\hbar
    &\approx&  \delta \hat{a}^\dagger\hat{a} + \tilde{\delta}_m \sum_i \hat{\sigma}_i^z \\
      &+& \sum_i \lambda_i \left\{ ( g_{R,1} \hat{a} + g_{R,2}\hat{a}^\dagger )\ket{\down}_i\!\bra{\up} + \hc \right\}, \nn
\eea 
where the two-photon couplings are given by $\lambda_i g_{R,1}=\Omega_1 \lambda_ig_0/2(\Delta+\Delta_F)$ and $\lambda_i g_{R,2}=\Omega_2 \lambda_i g_0/2\Delta$, respectively, 
$\hat{\sigma} ^z$ is the Pauli-z operator, and $\tilde{\delta}_m$ describes the spitting between the ground states AC-Stark shifted by the coupling beams.
By setting $\Omega_1/\Omega_2=(\Delta+\Delta_F)/\Delta$, 
we achieve equal coupling strengths $g_{R,1}=g_{R,2} \equiv g_R $.

When $|g_R/\delta|\ll 1$, both two-photon processes are suppressed. The two Raman processes have opposite detunings $\pm \delta$ since one of them absorbs a cavity photon while the other emits one.
Therefore, the four-photon process connecting two Raman couplings becomes resonant. 
As shown in Fig.~\ref{fig:experiment}(c), the total four processes realize the full $\lambda_i\lambda_j\hat{\sigma}_x^i\hat{\sigma}_x^j$ coupling.
For example, at the leftmost of Fig.~\ref{fig:experiment}(c), a Raman process $\lambda_i g_{R,1}\hat\sigma_i^{+}\hat a^{\dagger}$ of the $i$-th atom and a $\lambda_j g_{R,1}^*\hat\sigma_j^{-}\hat a$ process of the $j$-th atom can form a resonant process $\lambda_i g_{R,1}\hat \sigma^{+}\lambda_j g_{R,1}^*\hat \sigma^{-}$.
Even when the cavity is in vacuum states, the spins can still interact with each other via exchanging virtual photons through the cavity. 
Together with the other three similar processes that conserve the cavity photon number (see Fig.~\ref{fig:experiment}(c)), the system could realize the terms of $\hat \sigma^+_i \hat \sigma^+_j ,\hat \sigma^-_i \hat \sigma^-_j ,\hat \sigma^+_i \hat \sigma^-_j $ and $\hat \sigma^-_i \hat \sigma^+_j $. 
Since these four processes have equal coupling strengths, this leads to an effective Hamiltonian, 
\begin{equation}
    \begin{split}
        \hat H_{\text{eff}}/\hbar &= \sum_{i=1}^{N} \tilde{\delta}_m \hat \sigma_{i}^{z} + g_4\left(\sum_{i=1}^{N}\lambda_i\hat \sigma_i^x\right)^2 
    \end{split},
\label{eq:cavityeff} 
\end{equation}
Both 
$\tilde{\delta}_m$ 
and $g_4$ are dynamically tunable in experiments with constraints $|\Omega_{1, 2}| \ll  |\Delta|$, and  $|g_R|, |\tilde{\delta}_m| \ll |\delta|$.

To reach the many-body ground state of $\hat H_{\rm NPP}$, we apply the AQC protocol in the following way. First, we begin with a finite detuning $\tilde{\delta}_m(t=0)$ and set $g_4(t=0)=0$ by turning off the coupling beams $\Omega_1$ and $\Omega_2$. The ground state at this point is a trivial product state $\ket{\downarrow} ^{\otimes n}$ that can be easily prepared by optical pumping with high fidelity. 
Then we adiabatically ramp the Hamiltonian following a path $s(t)$ that goes from zero to unity.
That is, we control the detuning as $\tilde{\delta}_m(t) = [1-s(t)]\tilde{\delta}_m(0)$. Simultaneously, we ramp the intensity of the two incident lasers to the desired value as $\Omega_{1(2)}(t) = \sqrt{s(t)}\Omega_{1(2)}$, so that $g_4(t) = s(t)g_4$.
Finally, the solution to the NPP problem is obtained by measuring the atomic spins in the Pauli-$\hat{\sigma} ^ x$ basis.
There are several ways to improve the success probability of AQC, for example with optimizing the Hamiltonian path~\cite{lin2020quantum,lin2021hard,2022_Cirac_VQAA,2022_Hsieh_NMI} or by iterative reverse annealing schemes~\cite{2011_Aspuru_RQC,2017_Chancellor_NJP,2019_Lidar_RQA}.

For solving NPPs, besides our atom cavity proposal, there are also other candidate systems or protocols such as trapped ions~\cite{2016_Lewenstein_NC}, and Grover search in a central spin setup~\cite{anikeeva2021number}.
{Our proposing atom cavity realization is expected to be more scalable for its convenience in increasing atom numbers.}  
The theory we shall develop below may well be used to show all these systems have potential to solve generic quantum optimization with arbitrary connectivity. 
Nonetheless, we mainly consider  the atom cavity system in this work.  

\paragraph*{Encoding for 3-SAT.}

We consider  a 3-SAT problem with $n$ variables $x_{1},\dots,x_{n}$, and $m$ clauses $f_{1},\dots,f_{m}$~\cite{Kitaev_QCBook}. 
The 3-SAT problem is defined by two $m\times 3$ matrices $I$ and $B$, 
with 
\be 
f_j = ( x_{I_{j,1}}\oplus B_{j,1}) 
\lor ( x_{I_{j,2}}\oplus B_{j,2} ) 
\lor  ( x_{I_{j,3}}\oplus B_{j,3}) 
\label{eq:3-SATdef} 
\ee  
where $I$ contains integers from $1$ to $n$, and $B$ contains binary values. 
The idea is to map the 3-SAT problem to the atom cavity Hamiltonian in Eq.~\eqref{eq:npphamiltonian}.  
We introduce 
\bea 
    && \textstyle a_{i}=\sqrt{\alpha_{m+i}}+\sum_{j:x_{i} {\rm in} f_{j}}^{m}\sqrt{\alpha_{j}} \nn \\
    && \textstyle b_{i}=\sqrt{\alpha_{m+i}}+\sum_{j:\overline{x_{i}} {\rm in} f_{j}}^{m}\sqrt{\alpha_{j}},   \\ 
    && \textstyle c_j = d_j = \sqrt{\alpha_j} \nn
\eea 
with $\alpha_p$ the $p$-th squarefree integer, starting from $1$. Here, we have  $i\in [1, n]$, and $j \in [1, m]$. The numbers $\{a_i\}$, $\{b_i\}$, $\{c_j\}$, and $\{d_j\}$ form a set ${\cal R}$, which contains $2n+2m$ numbers,  to be referred to  as $r_k (I, B) $, with $k \in [1, 2n+2m]$. 
Solving 3-SAT problem is equivalent to finding a subset of ${\cal R}$ with its numbers added up to a target, 
\be 
   \textstyle  T(n,m) =\sum_{i=1}^{n}\sqrt{\alpha_{m+i}}+3\sum_{j=1}^{m}\sqrt{\alpha_{j}}, 
\ee 
or more specifically solving 
$
\sum_k y_k r_k (I, B) = T(n,m)
$
with $y_k$ taking binary values ($0$ or $1$). 
This equivalence is guaranteed by the property of  linear independence of radicals obeyed by the squarefree integers~\cite{2007_lam_mathematics}  (Supplementary Materials).

The 3-SAT problem  now becomes   optimizing 
\be 
\textstyle {\rm min}_{\{ y_k \} } \left( \sum_k y_k r_k (I, B)  -T  (n,m) \right)^2, 
\label{eq:3SATencoding} 
\ee 
which directly maps onto the atom cavity system (Eq.~\eqref{eq:npphamiltonian}). 
The optimization here only involves local fields and interactions of the factorized form of $r_{k} r_{k'}$ as existent in the cavity system. 
The additional requirement for encoding 3-SAT compared to NPP is the control over Rabi frequency for each atom, which is feasible to the atom cavity system through local addressing for each atom.

The ground state $y_{k} ^\star$ can be obtained by performing AQC. 
With our scheme, a 3-SAT with $n$ variables and $m$ clauses would cost $2n+2m$ atoms. The corresponding atom number overhead in this encoding is 
\be 
\textstyle {\rm Overhead} = n+2m. 
\ee 
The 3-SAT solution is directly given by $x_i = y_i ^\star $. 

We remark here that vertex cover problems can also be encoded by the atom cavity system in a similar way as 3-SAT (Supplementary Materials). 

We also mention that in computation theory, an alternative approach has been developed to construct the equivalence between NPP and 3-SAT, by using a series of integer-power of fours ($4^p$)~\cite{cormen2022introduction}, instead of squarefree integers. This would then require the atom couplings to grow exponentially with the atom number or to be exponentially precise, which is problematic for experimental implementation of large scale computation. This problem is absent in our construction using squarefree integers. 

\paragraph*{Encoding for QUBO.} 
QUBO is to minimize a quadratic objective function of binary variables with no constraints~\cite{2014_Wang_QUBO}. It corresponds to finding the ground state of an Ising model of $n$ classical spins ($s_i = \pm$) on a graph~\cite{2002_Roberto_Science}, 
\be 
    E(\{s_i\} )= \sum_{ i,i'>i}J_{ii'}s_{i}s_{i'} 
\ee 
Mapping this Ising model on a physical system requires engineering non-local interactions, which is challenging to implement in experiments. 
Although the atom cavity system has long-range interactions ($\lambda_i \lambda_{i'}$ in Eq.~\eqref{eq:cavityeff}), their form does not necessarily match $J_{ii'}$. The number of free parameters in $J$ scales quadratically with the number of Ising spins, whereas it scales linearly with the atom number  for the interactions of the atom cavity system. This implies the minimal number of encoding atoms has to scale quadratically with $n$.

We now develop a protocol for mapping the QUBO problem to the atom cavity system. To proceed, we first adopt the parity encoding, introducing $b_{ii'} = s_i s_{i'} $~\cite{2015_Zoller_SciAdv}. Treating $(ii')$ as a one single site of a square lattice (with size $n\times n$), the interactions in $J$ then become local fields, with a cost of introducing constraints 
$b_{ii'} b_{ii'+1} b_{i+1i'+1} b_{i+1i'} = 1$~\cite{2015_Zoller_SciAdv}.  The  total number of these independent constraints is $(n-1)(n-2)/2$. 
We then rewrite the constraints in terms of 3-SAT formula, 
\bea 
\label{eq:constraint} 
&& (\beta \lor x_{ii'} \lor x_{ii'+1})
	\land(\beta \lor\bar{x}_{ii'}\lor\bar{x}_{ii'+1}) \nn \\ 
&\land& (\beta \lor x _{i+1i'+1}\lor x_{i+1i'}) 
	\land(\beta \lor\bar{x}_{i+1i'+1}\lor\bar{x}_{i+1i'}) \nn \\ 
&\land&(\bar{\beta} \lor\bar{x}_{ii'}\lor x_{ii'+1}) 
	\land(\bar{\beta} \lor x_{ii'}\lor\bar{x}_{ii'+1})\\
&\land&(\bar{\beta} \lor\bar{x}_{i+1i'+1}\lor x_{i+1i'})
	\land(\bar{\beta} \lor x_{i+1i'+1}\lor\bar{x}_{i+1i'}), \nn 
\eea 
with $x_{ii'} = (b_{ii'}+1)/2$, and $\beta$ an introduced auxiliary variable. Taking all constraints into account, we have a 3-SAT problem with $n'= (n-1)^2$ variables, and $m'=4(n-1)(n-2)$ clauses. The corresponding defining matrices $I_{\rm const} $ and $B_{\rm const}$ are directly given according to Eqs.~\eqref{eq:3-SATdef} and \eqref{eq:constraint}. 

As shown above, the 3-SAT formula are equivalent to the optimization problem in Eq.~\eqref{eq:3SATencoding}. 
Now, the QUBO problem becomes optimizing 
\bea 
\label{eq:QUBOencoding} 
&& {\rm min}_{} \left\{
\sum_{k=1}^{n(n-1)/2} J_{k} y_k  \right. \\ 
&\quad& \left.  + {\cal M} \left(  \sum_{k=1} ^{2(n'+m') }  y_k r_k (I_{\rm const}, B_{\rm const} )- T(n',m')  \right) ^2 
			\right\}  \nn 
\eea  
with the first $n(n-1)/2$ elements of $y_k$ representing $x_{ii'}$, and $J_k$ representing $J_{ii'}$ correspondingly. 
Here, energy penalty term ${\cal M}  >0$ should be sufficiently large to enforce the required constraints. 
For practical AQC, it is suggested to start from  a moderate ${\cal M} $, and check for convergence upon its increase.

 Since the optimization in Eq.~\eqref{eq:QUBOencoding} only involves local fields and interactions of the factorized form of $r_{k} r_{k'}$, it maps directly to the atom cavity system. 
The solution of QUBO can be efficiently decoded from $y_k$ using the algorithm developed for parity encoding~\cite{2016_Preskill_ParityDecoder}. 
The overhead in the cost of atom number for the QUBO problem is 
\be 
\textstyle {\rm Overhead}  = 2(n-1)(5n-9)-n. 
\ee 
The quadratic scaling of the overhead is already optimal. We thus have a universal atom cavity based quantum optimization solver for generic QUBO problem, with the scaling of the cost of atom numbers being optimal. 
We remark here that the atom cavity solver for QUBO problem does not require arranging the atoms in a regular two-dimensional array. Our scheme fully exploits the form of the non-local interactions of the cavity system. 
 
\paragraph*{Discussion.---}

We develop a universal quantum optimization architecture based on cold atoms in an optical cavity. A Raman scheme is constructed using four-photon processes, where the cavity photon induced atomic interactions naturally encode NPPs. 
We find 3-SAT and vertex cover problems can also be efficiently encoded by the atom cavity system, at a linear cost of atom number, which is in contrast to the quadratic overhead in the Rydberg encoding~\cite{2020_Qiu_PRXQ, 2023_Pichler_PRXQ}.  
Based on the encoding scheme for 3-SAT, we further design an atom cavity architecture for generic QUBO problems with arbitrary connectivity. The atom number overhead for encoding QUBO has quadratic scaling, which is optimal for QUBO. 
We anticipate the atom cavity system to provide a compelling platform for quantum optimization having wide applications in academia and industry~\cite{hauke2020perspectives}. 
 With our theory, the cavity system has a potential to become a universal quantum optimization platform to demonstrate practical quantum advantage. 
The decoherence of this system  is mainly from the finite atomic excited state lifetime and the photon leakage, both of which are controllable by the two-photon detuning. Their tradeoff sets the limit of the quantum coherence time $T$. 
A worst-scenario estimate gives $g_4 T\propto \eta^{1/2}/N$, with $\eta$ the cavity cooperativity (Supplementary Material). 
The quantum coherence can be further improved by considering more advanced techniques such as cross cavities, by which we have 
 $g_4 T \propto \eta^{3/2}$ in the large atom number limit. 
 This implies that our scheme is scalable, and that 
 the coherence time can be systematically improved by advancing the cavity engineering technology.  

\paragraph*{Acknowledgement.} 
We acknowledge helpful discussion with Changling Zou and Lei Shi. 
This work is supported by 
National Key Research and Development Program of China (2021YFA1400900 and 2021YFA0718303),
National Natural Science Foundation of China (11934002, 92165203, 61975092, and 11974202), 
and Shanghai Municipal Science and Technology Major Project (Grant No. 2019SHZDZX01).

$^\S${These authors contributed equally to this work.} 

\bibliography{references}

\begin{thebibliography}{45}%
\makeatletter
\providecommand \@ifxundefined [1]{%
 \@ifx{#1\undefined}
}%
\providecommand \@ifnum [1]{%
 \ifnum #1\expandafter \@firstoftwo
 \else \expandafter \@secondoftwo
 \fi
}%
\providecommand \@ifx [1]{%
 \ifx #1\expandafter \@firstoftwo
 \else \expandafter \@secondoftwo
 \fi
}%
\providecommand \natexlab [1]{#1}%
\providecommand \enquote  [1]{``#1''}%
\providecommand \bibnamefont  [1]{#1}%
\providecommand \bibfnamefont [1]{#1}%
\providecommand \citenamefont [1]{#1}%
\providecommand \href@noop [0]{\@secondoftwo}%
\providecommand \href [0]{\begingroup \@sanitize@url \@href}%
\providecommand \@href[1]{\@@startlink{#1}\@@href}%
\providecommand \@@href[1]{\endgroup#1\@@endlink}%
\providecommand \@sanitize@url [0]{\catcode `\\12\catcode `\$12\catcode
  `\&12\catcode `\#12\catcode `\^12\catcode `\_12\catcode `\%12\relax}%
\providecommand \@@startlink[1]{}%
\providecommand \@@endlink[0]{}%
\providecommand \url  [0]{\begingroup\@sanitize@url \@url }%
\providecommand \@url [1]{\endgroup\@href {#1}{\urlprefix }}%
\providecommand \urlprefix  [0]{URL }%
\providecommand \Eprint [0]{\href }%
\providecommand \doibase [0]{http://dx.doi.org/}%
\providecommand \selectlanguage [0]{\@gobble}%
\providecommand \bibinfo  [0]{\@secondoftwo}%
\providecommand \bibfield  [0]{\@secondoftwo}%
\providecommand \translation [1]{[#1]}%
\providecommand \BibitemOpen [0]{}%
\providecommand \bibitemStop [0]{}%
\providecommand \bibitemNoStop [0]{.\EOS\space}%
\providecommand \EOS [0]{\spacefactor3000\relax}%
\providecommand \BibitemShut  [1]{\csname bibitem#1\endcsname}%
\let\auto@bib@innerbib\@empty
\bibitem [{\citenamefont {Farhi}\ \emph {et~al.}(2001)\citenamefont {Farhi},
  \citenamefont {Goldstone}, \citenamefont {Gutmann}, \citenamefont {Lapan},
  \citenamefont {Lundgren},\ and\ \citenamefont {Preda}}]{2001_Farhi_Science}%
  \BibitemOpen
  \bibfield  {author} {\bibinfo {author} {\bibfnamefont {E.}~\bibnamefont
  {Farhi}}, \bibinfo {author} {\bibfnamefont {J.}~\bibnamefont {Goldstone}},
  \bibinfo {author} {\bibfnamefont {S.}~\bibnamefont {Gutmann}}, \bibinfo
  {author} {\bibfnamefont {J.}~\bibnamefont {Lapan}}, \bibinfo {author}
  {\bibfnamefont {A.}~\bibnamefont {Lundgren}}, \ and\ \bibinfo {author}
  {\bibfnamefont {D.}~\bibnamefont {Preda}},\ }\href {\doibase
  10.1126/science.1057726} {\bibfield  {journal} {\bibinfo  {journal}
  {Science}\ }\textbf {\bibinfo {volume} {292}},\ \bibinfo {pages} {472}
  (\bibinfo {year} {2001})}\BibitemShut {NoStop}%
\bibitem [{\citenamefont {Farhi}\ \emph {et~al.}(2014)\citenamefont {Farhi},
  \citenamefont {Goldstone},\ and\ \citenamefont {Gutmann}}]{2014_Farhi_QAOA}%
  \BibitemOpen
  \bibfield  {author} {\bibinfo {author} {\bibfnamefont {E.}~\bibnamefont
  {Farhi}}, \bibinfo {author} {\bibfnamefont {J.}~\bibnamefont {Goldstone}}, \
  and\ \bibinfo {author} {\bibfnamefont {S.}~\bibnamefont {Gutmann}},\
  }\href@noop {} {\bibfield  {journal} {\bibinfo  {journal} {arXiv preprint
  arXiv:1411.4028}\ } (\bibinfo {year} {2014})}\BibitemShut {NoStop}%
\bibitem [{\citenamefont {Perdomo-Ortiz}\ \emph {et~al.}(2012)\citenamefont
  {Perdomo-Ortiz}, \citenamefont {Dickson}, \citenamefont {Drew-Brook},
  \citenamefont {Rose},\ and\ \citenamefont
  {Aspuru-Guzik}}]{2012_Aspuru-Guzik_SciRep}%
  \BibitemOpen
  \bibfield  {author} {\bibinfo {author} {\bibfnamefont {A.}~\bibnamefont
  {Perdomo-Ortiz}}, \bibinfo {author} {\bibfnamefont {N.}~\bibnamefont
  {Dickson}}, \bibinfo {author} {\bibfnamefont {M.}~\bibnamefont {Drew-Brook}},
  \bibinfo {author} {\bibfnamefont {G.}~\bibnamefont {Rose}}, \ and\ \bibinfo
  {author} {\bibfnamefont {A.}~\bibnamefont {Aspuru-Guzik}},\ }\href {\doibase
  10.1038/srep00571} {\bibfield  {journal} {\bibinfo  {journal} {Scientific
  Reports}\ }\textbf {\bibinfo {volume} {2}},\ \bibinfo {pages} {571} (\bibinfo
  {year} {2012})}\BibitemShut {NoStop}%
\bibitem [{\citenamefont {Gaidai}\ \emph {et~al.}(2022)\citenamefont {Gaidai},
  \citenamefont {Babikov}, \citenamefont {Teplukhin}, \citenamefont {Kendrick},
  \citenamefont {Mniszewski}, \citenamefont {Zhang}, \citenamefont {Tretiak},\
  and\ \citenamefont {Dub}}]{2022_Gaidai_MD}%
  \BibitemOpen
  \bibfield  {author} {\bibinfo {author} {\bibfnamefont {I.}~\bibnamefont
  {Gaidai}}, \bibinfo {author} {\bibfnamefont {D.}~\bibnamefont {Babikov}},
  \bibinfo {author} {\bibfnamefont {A.}~\bibnamefont {Teplukhin}}, \bibinfo
  {author} {\bibfnamefont {B.~K.}\ \bibnamefont {Kendrick}}, \bibinfo {author}
  {\bibfnamefont {S.~M.}\ \bibnamefont {Mniszewski}}, \bibinfo {author}
  {\bibfnamefont {Y.}~\bibnamefont {Zhang}}, \bibinfo {author} {\bibfnamefont
  {S.}~\bibnamefont {Tretiak}}, \ and\ \bibinfo {author} {\bibfnamefont
  {P.~A.}\ \bibnamefont {Dub}},\ }\href {\doibase 10.1038/s41598-022-21163-x}
  {\bibfield  {journal} {\bibinfo  {journal} {Scientific Reports}\ }\textbf
  {\bibinfo {volume} {12}},\ \bibinfo {pages} {16824} (\bibinfo {year}
  {2022})}\BibitemShut {NoStop}%
\bibitem [{\citenamefont {Clark}\ \emph {et~al.}(1990)\citenamefont {Clark},
  \citenamefont {Colbourn},\ and\ \citenamefont {Johnson}}]{1990_Unit_disk}%
  \BibitemOpen
  \bibfield  {author} {\bibinfo {author} {\bibfnamefont {B.~N.}\ \bibnamefont
  {Clark}}, \bibinfo {author} {\bibfnamefont {C.~J.}\ \bibnamefont {Colbourn}},
  \ and\ \bibinfo {author} {\bibfnamefont {D.~S.}\ \bibnamefont {Johnson}},\
  }\href@noop {} {\bibfield  {journal} {\bibinfo  {journal} {Discrete
  mathematics}\ }\textbf {\bibinfo {volume} {86}},\ \bibinfo {pages} {165}
  (\bibinfo {year} {1990})}\BibitemShut {NoStop}%
\bibitem [{\citenamefont {Zhou}\ \emph {et~al.}(2020)\citenamefont {Zhou},
  \citenamefont {Wang}, \citenamefont {Choi}, \citenamefont {Pichler},\ and\
  \citenamefont {Lukin}}]{2020_Zhou_PRX}%
  \BibitemOpen
  \bibfield  {author} {\bibinfo {author} {\bibfnamefont {L.}~\bibnamefont
  {Zhou}}, \bibinfo {author} {\bibfnamefont {S.-T.}\ \bibnamefont {Wang}},
  \bibinfo {author} {\bibfnamefont {S.}~\bibnamefont {Choi}}, \bibinfo {author}
  {\bibfnamefont {H.}~\bibnamefont {Pichler}}, \ and\ \bibinfo {author}
  {\bibfnamefont {M.~D.}\ \bibnamefont {Lukin}},\ }\href {\doibase
  10.1103/PhysRevX.10.021067} {\bibfield  {journal} {\bibinfo  {journal} {Phys.
  Rev. X}\ }\textbf {\bibinfo {volume} {10}},\ \bibinfo {pages} {021067}
  (\bibinfo {year} {2020})}\BibitemShut {NoStop}%
\bibitem [{\citenamefont {Hastings}(2021)}]{2021_Hastings_Quantum}%
  \BibitemOpen
  \bibfield  {author} {\bibinfo {author} {\bibfnamefont {M.~B.}\ \bibnamefont
  {Hastings}},\ }\href@noop {} {\bibfield  {journal} {\bibinfo  {journal}
  {Quantum}\ }\textbf {\bibinfo {volume} {5}},\ \bibinfo {pages} {597}
  (\bibinfo {year} {2021})}\BibitemShut {NoStop}%
\bibitem [{\citenamefont {Ebadi}\ \emph {et~al.}(2022)\citenamefont {Ebadi},
  \citenamefont {Keesling}, \citenamefont {Cain}, \citenamefont {Wang},
  \citenamefont {Levine}, \citenamefont {Bluvstein}, \citenamefont {Semeghini},
  \citenamefont {Omran}, \citenamefont {Liu}, \citenamefont {Samajdar},
  \citenamefont {Luo}, \citenamefont {Nash}, \citenamefont {Gao}, \citenamefont
  {Barak}, \citenamefont {Farhi}, \citenamefont {Sachdev}, \citenamefont
  {Gemelke}, \citenamefont {Zhou}, \citenamefont {Choi}, \citenamefont
  {Pichler}, \citenamefont {Wang}, \citenamefont {Greiner}, \citenamefont
  {Vuleti{\'c}},\ and\ \citenamefont {Lukin}}]{2022_Lukin_Science}%
  \BibitemOpen
  \bibfield  {author} {\bibinfo {author} {\bibfnamefont {S.}~\bibnamefont
  {Ebadi}}, \bibinfo {author} {\bibfnamefont {A.}~\bibnamefont {Keesling}},
  \bibinfo {author} {\bibfnamefont {M.}~\bibnamefont {Cain}}, \bibinfo {author}
  {\bibfnamefont {T.~T.}\ \bibnamefont {Wang}}, \bibinfo {author}
  {\bibfnamefont {H.}~\bibnamefont {Levine}}, \bibinfo {author} {\bibfnamefont
  {D.}~\bibnamefont {Bluvstein}}, \bibinfo {author} {\bibfnamefont
  {G.}~\bibnamefont {Semeghini}}, \bibinfo {author} {\bibfnamefont
  {A.}~\bibnamefont {Omran}}, \bibinfo {author} {\bibfnamefont {J.-G.}\
  \bibnamefont {Liu}}, \bibinfo {author} {\bibfnamefont {R.}~\bibnamefont
  {Samajdar}}, \bibinfo {author} {\bibfnamefont {X.-Z.}\ \bibnamefont {Luo}},
  \bibinfo {author} {\bibfnamefont {B.}~\bibnamefont {Nash}}, \bibinfo {author}
  {\bibfnamefont {X.}~\bibnamefont {Gao}}, \bibinfo {author} {\bibfnamefont
  {B.}~\bibnamefont {Barak}}, \bibinfo {author} {\bibfnamefont
  {E.}~\bibnamefont {Farhi}}, \bibinfo {author} {\bibfnamefont
  {S.}~\bibnamefont {Sachdev}}, \bibinfo {author} {\bibfnamefont
  {N.}~\bibnamefont {Gemelke}}, \bibinfo {author} {\bibfnamefont
  {L.}~\bibnamefont {Zhou}}, \bibinfo {author} {\bibfnamefont {S.}~\bibnamefont
  {Choi}}, \bibinfo {author} {\bibfnamefont {H.}~\bibnamefont {Pichler}},
  \bibinfo {author} {\bibfnamefont {S.-T.}\ \bibnamefont {Wang}}, \bibinfo
  {author} {\bibfnamefont {M.}~\bibnamefont {Greiner}}, \bibinfo {author}
  {\bibfnamefont {V.}~\bibnamefont {Vuleti{\'c}}}, \ and\ \bibinfo {author}
  {\bibfnamefont {M.~D.}\ \bibnamefont {Lukin}},\ }\href {\doibase
  10.1126/science.abo6587} {\bibfield  {journal} {\bibinfo  {journal}
  {Science}\ }\textbf {\bibinfo {volume} {376}},\ \bibinfo {pages} {1209}
  (\bibinfo {year} {2022})}\BibitemShut {NoStop}%
\bibitem [{\citenamefont {King}\ \emph {et~al.}(2023)\citenamefont {King},
  \citenamefont {Raymond}, \citenamefont {Lanting}, \citenamefont {Harris},
  \citenamefont {Zucca}, \citenamefont {Altomare}, \citenamefont {Berkley},
  \citenamefont {Boothby}, \citenamefont {Ejtemaee}, \citenamefont {Enderud},
  \citenamefont {Hoskinson}, \citenamefont {Huang}, \citenamefont {Ladizinsky},
  \citenamefont {MacDonald}, \citenamefont {Marsden}, \citenamefont {Molavi},
  \citenamefont {Oh}, \citenamefont {Poulin-Lamarre}, \citenamefont {Reis},
  \citenamefont {Rich}, \citenamefont {Sato}, \citenamefont {Tsai},
  \citenamefont {Volkmann}, \citenamefont {Whittaker}, \citenamefont {Yao},
  \citenamefont {Sandvik},\ and\ \citenamefont {Amin}}]{2023_Dwave_Nature}%
  \BibitemOpen
  \bibfield  {author} {\bibinfo {author} {\bibfnamefont {A.~D.}\ \bibnamefont
  {King}}, \bibinfo {author} {\bibfnamefont {J.}~\bibnamefont {Raymond}},
  \bibinfo {author} {\bibfnamefont {T.}~\bibnamefont {Lanting}}, \bibinfo
  {author} {\bibfnamefont {R.}~\bibnamefont {Harris}}, \bibinfo {author}
  {\bibfnamefont {A.}~\bibnamefont {Zucca}}, \bibinfo {author} {\bibfnamefont
  {F.}~\bibnamefont {Altomare}}, \bibinfo {author} {\bibfnamefont {A.~J.}\
  \bibnamefont {Berkley}}, \bibinfo {author} {\bibfnamefont {K.}~\bibnamefont
  {Boothby}}, \bibinfo {author} {\bibfnamefont {S.}~\bibnamefont {Ejtemaee}},
  \bibinfo {author} {\bibfnamefont {C.}~\bibnamefont {Enderud}}, \bibinfo
  {author} {\bibfnamefont {E.}~\bibnamefont {Hoskinson}}, \bibinfo {author}
  {\bibfnamefont {S.}~\bibnamefont {Huang}}, \bibinfo {author} {\bibfnamefont
  {E.}~\bibnamefont {Ladizinsky}}, \bibinfo {author} {\bibfnamefont {A.~J.~R.}\
  \bibnamefont {MacDonald}}, \bibinfo {author} {\bibfnamefont {G.}~\bibnamefont
  {Marsden}}, \bibinfo {author} {\bibfnamefont {R.}~\bibnamefont {Molavi}},
  \bibinfo {author} {\bibfnamefont {T.}~\bibnamefont {Oh}}, \bibinfo {author}
  {\bibfnamefont {G.}~\bibnamefont {Poulin-Lamarre}}, \bibinfo {author}
  {\bibfnamefont {M.}~\bibnamefont {Reis}}, \bibinfo {author} {\bibfnamefont
  {C.}~\bibnamefont {Rich}}, \bibinfo {author} {\bibfnamefont {Y.}~\bibnamefont
  {Sato}}, \bibinfo {author} {\bibfnamefont {N.}~\bibnamefont {Tsai}}, \bibinfo
  {author} {\bibfnamefont {M.}~\bibnamefont {Volkmann}}, \bibinfo {author}
  {\bibfnamefont {J.~D.}\ \bibnamefont {Whittaker}}, \bibinfo {author}
  {\bibfnamefont {J.}~\bibnamefont {Yao}}, \bibinfo {author} {\bibfnamefont
  {A.~W.}\ \bibnamefont {Sandvik}}, \ and\ \bibinfo {author} {\bibfnamefont
  {M.~H.}\ \bibnamefont {Amin}},\ }\href {\doibase 10.1038/s41586-023-05867-2}
  {\bibfield  {journal} {\bibinfo  {journal} {Nature}\ }\textbf {\bibinfo
  {volume} {617}},\ \bibinfo {pages} {61} (\bibinfo {year} {2023})}\BibitemShut
  {NoStop}%
\bibitem [{\citenamefont {Lin}\ \emph {et~al.}(2020)\citenamefont {Lin},
  \citenamefont {Lai},\ and\ \citenamefont {Li}}]{lin2020quantum}%
  \BibitemOpen
  \bibfield  {author} {\bibinfo {author} {\bibfnamefont {J.}~\bibnamefont
  {Lin}}, \bibinfo {author} {\bibfnamefont {Z.~Y.}\ \bibnamefont {Lai}}, \ and\
  \bibinfo {author} {\bibfnamefont {X.}~\bibnamefont {Li}},\ }\href@noop {}
  {\bibfield  {journal} {\bibinfo  {journal} {Physical Review A}\ }\textbf
  {\bibinfo {volume} {101}},\ \bibinfo {pages} {052327} (\bibinfo {year}
  {2020})}\BibitemShut {NoStop}%
\bibitem [{\citenamefont {Lin}\ \emph {et~al.}(2021)\citenamefont {Lin},
  \citenamefont {Zhang}, \citenamefont {Zhang},\ and\ \citenamefont
  {Li}}]{lin2021hard}%
  \BibitemOpen
  \bibfield  {author} {\bibinfo {author} {\bibfnamefont {J.}~\bibnamefont
  {Lin}}, \bibinfo {author} {\bibfnamefont {Z.}~\bibnamefont {Zhang}}, \bibinfo
  {author} {\bibfnamefont {J.}~\bibnamefont {Zhang}}, \ and\ \bibinfo {author}
  {\bibfnamefont {X.}~\bibnamefont {Li}},\ }\href@noop {} {\bibfield  {journal}
  {\bibinfo  {journal} {arXiv preprint arXiv:2110.04782}\ } (\bibinfo {year}
  {2021})}\BibitemShut {NoStop}%
\bibitem [{\citenamefont {Schiffer}\ \emph {et~al.}(2022)\citenamefont
  {Schiffer}, \citenamefont {Tura},\ and\ \citenamefont
  {Cirac}}]{2022_Cirac_VQAA}%
  \BibitemOpen
  \bibfield  {author} {\bibinfo {author} {\bibfnamefont {B.~F.}\ \bibnamefont
  {Schiffer}}, \bibinfo {author} {\bibfnamefont {J.}~\bibnamefont {Tura}}, \
  and\ \bibinfo {author} {\bibfnamefont {J.~I.}\ \bibnamefont {Cirac}},\ }\href
  {\doibase 10.1103/PRXQuantum.3.020347} {\bibfield  {journal} {\bibinfo
  {journal} {PRX Quantum}\ }\textbf {\bibinfo {volume} {3}},\ \bibinfo {pages}
  {020347} (\bibinfo {year} {2022})}\BibitemShut {NoStop}%
\bibitem [{\citenamefont {Chen}\ \emph {et~al.}(2022)\citenamefont {Chen},
  \citenamefont {Chen}, \citenamefont {Lee}, \citenamefont {Zhang},\ and\
  \citenamefont {Hsieh}}]{2022_Hsieh_NMI}%
  \BibitemOpen
  \bibfield  {author} {\bibinfo {author} {\bibfnamefont {Y.-Q.}\ \bibnamefont
  {Chen}}, \bibinfo {author} {\bibfnamefont {Y.}~\bibnamefont {Chen}}, \bibinfo
  {author} {\bibfnamefont {C.-K.}\ \bibnamefont {Lee}}, \bibinfo {author}
  {\bibfnamefont {S.}~\bibnamefont {Zhang}}, \ and\ \bibinfo {author}
  {\bibfnamefont {C.-Y.}\ \bibnamefont {Hsieh}},\ }\href@noop {} {\bibfield
  {journal} {\bibinfo  {journal} {Nature Machine Intelligence}\ }\textbf
  {\bibinfo {volume} {4}},\ \bibinfo {pages} {269} (\bibinfo {year}
  {2022})}\BibitemShut {NoStop}%
\bibitem [{\citenamefont {Browaeys}\ and\ \citenamefont
  {Lahaye}(2020)}]{2020_Browaeys_NatPhys}%
  \BibitemOpen
  \bibfield  {author} {\bibinfo {author} {\bibfnamefont {A.}~\bibnamefont
  {Browaeys}}\ and\ \bibinfo {author} {\bibfnamefont {T.}~\bibnamefont
  {Lahaye}},\ }\href {\doibase 10.1038/s41567-019-0733-z} {\bibfield  {journal}
  {\bibinfo  {journal} {Nature Physics}\ }\textbf {\bibinfo {volume} {16}},\
  \bibinfo {pages} {132} (\bibinfo {year} {2020})}\BibitemShut {NoStop}%
\bibitem [{\citenamefont {Qiu}\ \emph {et~al.}(2020)\citenamefont {Qiu},
  \citenamefont {Zoller},\ and\ \citenamefont {Li}}]{2020_Qiu_PRXQ}%
  \BibitemOpen
  \bibfield  {author} {\bibinfo {author} {\bibfnamefont {X.}~\bibnamefont
  {Qiu}}, \bibinfo {author} {\bibfnamefont {P.}~\bibnamefont {Zoller}}, \ and\
  \bibinfo {author} {\bibfnamefont {X.}~\bibnamefont {Li}},\ }\href {\doibase
  10.1103/PRXQuantum.1.020311} {\bibfield  {journal} {\bibinfo  {journal} {PRX
  Quantum}\ }\textbf {\bibinfo {volume} {1}},\ \bibinfo {pages} {020311}
  (\bibinfo {year} {2020})}\BibitemShut {NoStop}%
\bibitem [{\citenamefont {Nguyen}\ \emph {et~al.}(2023)\citenamefont {Nguyen},
  \citenamefont {Liu}, \citenamefont {Wurtz}, \citenamefont {Lukin},
  \citenamefont {Wang},\ and\ \citenamefont {Pichler}}]{2023_Pichler_PRXQ}%
  \BibitemOpen
  \bibfield  {author} {\bibinfo {author} {\bibfnamefont {M.-T.}\ \bibnamefont
  {Nguyen}}, \bibinfo {author} {\bibfnamefont {J.-G.}\ \bibnamefont {Liu}},
  \bibinfo {author} {\bibfnamefont {J.}~\bibnamefont {Wurtz}}, \bibinfo
  {author} {\bibfnamefont {M.~D.}\ \bibnamefont {Lukin}}, \bibinfo {author}
  {\bibfnamefont {S.-T.}\ \bibnamefont {Wang}}, \ and\ \bibinfo {author}
  {\bibfnamefont {H.}~\bibnamefont {Pichler}},\ }\href {\doibase
  10.1103/PRXQuantum.4.010316} {\bibfield  {journal} {\bibinfo  {journal} {PRX
  Quantum}\ }\textbf {\bibinfo {volume} {4}},\ \bibinfo {pages} {010316}
  (\bibinfo {year} {2023})}\BibitemShut {NoStop}%
\bibitem [{\citenamefont {Scholl}\ \emph {et~al.}(2021)\citenamefont {Scholl},
  \citenamefont {Schuler}, \citenamefont {Williams}, \citenamefont
  {Eberharter}, \citenamefont {Barredo}, \citenamefont {Schymik}, \citenamefont
  {Lienhard}, \citenamefont {Henry}, \citenamefont {Lang}, \citenamefont
  {Lahaye}, \citenamefont {L{\"a}uchli},\ and\ \citenamefont
  {Browaeys}}]{2021_Browaeys_Nature}%
  \BibitemOpen
  \bibfield  {author} {\bibinfo {author} {\bibfnamefont {P.}~\bibnamefont
  {Scholl}}, \bibinfo {author} {\bibfnamefont {M.}~\bibnamefont {Schuler}},
  \bibinfo {author} {\bibfnamefont {H.~J.}\ \bibnamefont {Williams}}, \bibinfo
  {author} {\bibfnamefont {A.~A.}\ \bibnamefont {Eberharter}}, \bibinfo
  {author} {\bibfnamefont {D.}~\bibnamefont {Barredo}}, \bibinfo {author}
  {\bibfnamefont {K.-N.}\ \bibnamefont {Schymik}}, \bibinfo {author}
  {\bibfnamefont {V.}~\bibnamefont {Lienhard}}, \bibinfo {author}
  {\bibfnamefont {L.-P.}\ \bibnamefont {Henry}}, \bibinfo {author}
  {\bibfnamefont {T.~C.}\ \bibnamefont {Lang}}, \bibinfo {author}
  {\bibfnamefont {T.}~\bibnamefont {Lahaye}}, \bibinfo {author} {\bibfnamefont
  {A.~M.}\ \bibnamefont {L{\"a}uchli}}, \ and\ \bibinfo {author} {\bibfnamefont
  {A.}~\bibnamefont {Browaeys}},\ }\href {\doibase 10.1038/s41586-021-03585-1}
  {\bibfield  {journal} {\bibinfo  {journal} {Nature}\ }\textbf {\bibinfo
  {volume} {595}},\ \bibinfo {pages} {233} (\bibinfo {year}
  {2021})}\BibitemShut {NoStop}%
\bibitem [{\citenamefont {Ebadi}\ \emph {et~al.}(2021)\citenamefont {Ebadi},
  \citenamefont {Wang}, \citenamefont {Levine}, \citenamefont {Keesling},
  \citenamefont {Semeghini}, \citenamefont {Omran}, \citenamefont {Bluvstein},
  \citenamefont {Samajdar}, \citenamefont {Pichler}, \citenamefont {Ho},
  \citenamefont {Choi}, \citenamefont {Sachdev}, \citenamefont {Greiner},
  \citenamefont {Vuleti{\'c}},\ and\ \citenamefont
  {Lukin}}]{2021_Lukin_Rydberg}%
  \BibitemOpen
  \bibfield  {author} {\bibinfo {author} {\bibfnamefont {S.}~\bibnamefont
  {Ebadi}}, \bibinfo {author} {\bibfnamefont {T.~T.}\ \bibnamefont {Wang}},
  \bibinfo {author} {\bibfnamefont {H.}~\bibnamefont {Levine}}, \bibinfo
  {author} {\bibfnamefont {A.}~\bibnamefont {Keesling}}, \bibinfo {author}
  {\bibfnamefont {G.}~\bibnamefont {Semeghini}}, \bibinfo {author}
  {\bibfnamefont {A.}~\bibnamefont {Omran}}, \bibinfo {author} {\bibfnamefont
  {D.}~\bibnamefont {Bluvstein}}, \bibinfo {author} {\bibfnamefont
  {R.}~\bibnamefont {Samajdar}}, \bibinfo {author} {\bibfnamefont
  {H.}~\bibnamefont {Pichler}}, \bibinfo {author} {\bibfnamefont {W.~W.}\
  \bibnamefont {Ho}}, \bibinfo {author} {\bibfnamefont {S.}~\bibnamefont
  {Choi}}, \bibinfo {author} {\bibfnamefont {S.}~\bibnamefont {Sachdev}},
  \bibinfo {author} {\bibfnamefont {M.}~\bibnamefont {Greiner}}, \bibinfo
  {author} {\bibfnamefont {V.}~\bibnamefont {Vuleti{\'c}}}, \ and\ \bibinfo
  {author} {\bibfnamefont {M.~D.}\ \bibnamefont {Lukin}},\ }\href {\doibase
  10.1038/s41586-021-03582-4} {\bibfield  {journal} {\bibinfo  {journal}
  {Nature}\ }\textbf {\bibinfo {volume} {595}},\ \bibinfo {pages} {227}
  (\bibinfo {year} {2021})}\BibitemShut {NoStop}%
\bibitem [{\citenamefont {Kim}\ \emph {et~al.}(2022)\citenamefont {Kim},
  \citenamefont {Kim}, \citenamefont {Hwang}, \citenamefont {Moon},\ and\
  \citenamefont {Ahn}}]{2022_Ahn_NatPhys}%
  \BibitemOpen
  \bibfield  {author} {\bibinfo {author} {\bibfnamefont {M.}~\bibnamefont
  {Kim}}, \bibinfo {author} {\bibfnamefont {K.}~\bibnamefont {Kim}}, \bibinfo
  {author} {\bibfnamefont {J.}~\bibnamefont {Hwang}}, \bibinfo {author}
  {\bibfnamefont {E.-G.}\ \bibnamefont {Moon}}, \ and\ \bibinfo {author}
  {\bibfnamefont {J.}~\bibnamefont {Ahn}},\ }\href {\doibase
  10.1038/s41567-022-01629-5} {\bibfield  {journal} {\bibinfo  {journal}
  {Nature Physics}\ }\textbf {\bibinfo {volume} {18}},\ \bibinfo {pages} {755}
  (\bibinfo {year} {2022})}\BibitemShut {NoStop}%
\bibitem [{\citenamefont {Graham}\ \emph {et~al.}(2022)\citenamefont {Graham},
  \citenamefont {Song}, \citenamefont {Scott}, \citenamefont {Poole},
  \citenamefont {Phuttitarn}, \citenamefont {Jooya}, \citenamefont {Eichler},
  \citenamefont {Jiang}, \citenamefont {Marra}, \citenamefont {Grinkemeyer},
  \citenamefont {Kwon}, \citenamefont {Ebert}, \citenamefont {Cherek},
  \citenamefont {Lichtman}, \citenamefont {Gillette}, \citenamefont {Gilbert},
  \citenamefont {Bowman}, \citenamefont {Ballance}, \citenamefont {Campbell},
  \citenamefont {Dahl}, \citenamefont {Crawford}, \citenamefont {Blunt},
  \citenamefont {Rogers}, \citenamefont {Noel},\ and\ \citenamefont
  {Saffman}}]{2022_Saffman_Nature}%
  \BibitemOpen
  \bibfield  {author} {\bibinfo {author} {\bibfnamefont {T.~M.}\ \bibnamefont
  {Graham}}, \bibinfo {author} {\bibfnamefont {Y.}~\bibnamefont {Song}},
  \bibinfo {author} {\bibfnamefont {J.}~\bibnamefont {Scott}}, \bibinfo
  {author} {\bibfnamefont {C.}~\bibnamefont {Poole}}, \bibinfo {author}
  {\bibfnamefont {L.}~\bibnamefont {Phuttitarn}}, \bibinfo {author}
  {\bibfnamefont {K.}~\bibnamefont {Jooya}}, \bibinfo {author} {\bibfnamefont
  {P.}~\bibnamefont {Eichler}}, \bibinfo {author} {\bibfnamefont
  {X.}~\bibnamefont {Jiang}}, \bibinfo {author} {\bibfnamefont
  {A.}~\bibnamefont {Marra}}, \bibinfo {author} {\bibfnamefont
  {B.}~\bibnamefont {Grinkemeyer}}, \bibinfo {author} {\bibfnamefont
  {M.}~\bibnamefont {Kwon}}, \bibinfo {author} {\bibfnamefont {M.}~\bibnamefont
  {Ebert}}, \bibinfo {author} {\bibfnamefont {J.}~\bibnamefont {Cherek}},
  \bibinfo {author} {\bibfnamefont {M.~T.}\ \bibnamefont {Lichtman}}, \bibinfo
  {author} {\bibfnamefont {M.}~\bibnamefont {Gillette}}, \bibinfo {author}
  {\bibfnamefont {J.}~\bibnamefont {Gilbert}}, \bibinfo {author} {\bibfnamefont
  {D.}~\bibnamefont {Bowman}}, \bibinfo {author} {\bibfnamefont
  {T.}~\bibnamefont {Ballance}}, \bibinfo {author} {\bibfnamefont
  {C.}~\bibnamefont {Campbell}}, \bibinfo {author} {\bibfnamefont {E.~D.}\
  \bibnamefont {Dahl}}, \bibinfo {author} {\bibfnamefont {O.}~\bibnamefont
  {Crawford}}, \bibinfo {author} {\bibfnamefont {N.~S.}\ \bibnamefont {Blunt}},
  \bibinfo {author} {\bibfnamefont {B.}~\bibnamefont {Rogers}}, \bibinfo
  {author} {\bibfnamefont {T.}~\bibnamefont {Noel}}, \ and\ \bibinfo {author}
  {\bibfnamefont {M.}~\bibnamefont {Saffman}},\ }\href {\doibase
  10.1038/s41586-022-04603-6} {\bibfield  {journal} {\bibinfo  {journal}
  {Nature}\ }\textbf {\bibinfo {volume} {604}},\ \bibinfo {pages} {457}
  (\bibinfo {year} {2022})}\BibitemShut {NoStop}%
\bibitem [{\citenamefont {Sheng}\ \emph {et~al.}(2022)\citenamefont {Sheng},
  \citenamefont {Hou}, \citenamefont {He}, \citenamefont {Wang}, \citenamefont
  {Guo}, \citenamefont {Zhuang}, \citenamefont {Mamat}, \citenamefont {Xu},
  \citenamefont {Liu}, \citenamefont {Wang},\ and\ \citenamefont
  {Zhan}}]{2022_Sheng_PRL}%
  \BibitemOpen
  \bibfield  {author} {\bibinfo {author} {\bibfnamefont {C.}~\bibnamefont
  {Sheng}}, \bibinfo {author} {\bibfnamefont {J.}~\bibnamefont {Hou}}, \bibinfo
  {author} {\bibfnamefont {X.}~\bibnamefont {He}}, \bibinfo {author}
  {\bibfnamefont {K.}~\bibnamefont {Wang}}, \bibinfo {author} {\bibfnamefont
  {R.}~\bibnamefont {Guo}}, \bibinfo {author} {\bibfnamefont {J.}~\bibnamefont
  {Zhuang}}, \bibinfo {author} {\bibfnamefont {B.}~\bibnamefont {Mamat}},
  \bibinfo {author} {\bibfnamefont {P.}~\bibnamefont {Xu}}, \bibinfo {author}
  {\bibfnamefont {M.}~\bibnamefont {Liu}}, \bibinfo {author} {\bibfnamefont
  {J.}~\bibnamefont {Wang}}, \ and\ \bibinfo {author} {\bibfnamefont
  {M.}~\bibnamefont {Zhan}},\ }\href {\doibase 10.1103/PhysRevLett.128.083202}
  {\bibfield  {journal} {\bibinfo  {journal} {Phys. Rev. Lett.}\ }\textbf
  {\bibinfo {volume} {128}},\ \bibinfo {pages} {083202} (\bibinfo {year}
  {2022})}\BibitemShut {NoStop}%
\bibitem [{\citenamefont {Saffman}\ \emph {et~al.}(2010)\citenamefont
  {Saffman}, \citenamefont {Walker},\ and\ \citenamefont
  {M\o{}lmer}}]{2010_Saffman_RMP}%
  \BibitemOpen
  \bibfield  {author} {\bibinfo {author} {\bibfnamefont {M.}~\bibnamefont
  {Saffman}}, \bibinfo {author} {\bibfnamefont {T.~G.}\ \bibnamefont {Walker}},
  \ and\ \bibinfo {author} {\bibfnamefont {K.}~\bibnamefont {M\o{}lmer}},\
  }\href {\doibase 10.1103/RevModPhys.82.2313} {\bibfield  {journal} {\bibinfo
  {journal} {Rev. Mod. Phys.}\ }\textbf {\bibinfo {volume} {82}},\ \bibinfo
  {pages} {2313} (\bibinfo {year} {2010})}\BibitemShut {NoStop}%
\bibitem [{\citenamefont {Degen}\ \emph {et~al.}(2017)\citenamefont {Degen},
  \citenamefont {Reinhard},\ and\ \citenamefont
  {Cappellaro}}]{2017_Cappellaro_RMP}%
  \BibitemOpen
  \bibfield  {author} {\bibinfo {author} {\bibfnamefont {C.~L.}\ \bibnamefont
  {Degen}}, \bibinfo {author} {\bibfnamefont {F.}~\bibnamefont {Reinhard}}, \
  and\ \bibinfo {author} {\bibfnamefont {P.}~\bibnamefont {Cappellaro}},\
  }\href {\doibase 10.1103/RevModPhys.89.035002} {\bibfield  {journal}
  {\bibinfo  {journal} {Rev. Mod. Phys.}\ }\textbf {\bibinfo {volume} {89}},\
  \bibinfo {pages} {035002} (\bibinfo {year} {2017})}\BibitemShut {NoStop}%
\bibitem [{\citenamefont {Ritsch}\ \emph {et~al.}(2013)\citenamefont {Ritsch},
  \citenamefont {Domokos}, \citenamefont {Brennecke},\ and\ \citenamefont
  {Esslinger}}]{2013_Esslinger_RMP}%
  \BibitemOpen
  \bibfield  {author} {\bibinfo {author} {\bibfnamefont {H.}~\bibnamefont
  {Ritsch}}, \bibinfo {author} {\bibfnamefont {P.}~\bibnamefont {Domokos}},
  \bibinfo {author} {\bibfnamefont {F.}~\bibnamefont {Brennecke}}, \ and\
  \bibinfo {author} {\bibfnamefont {T.}~\bibnamefont {Esslinger}},\ }\href
  {\doibase 10.1103/RevModPhys.85.553} {\bibfield  {journal} {\bibinfo
  {journal} {Rev. Mod. Phys.}\ }\textbf {\bibinfo {volume} {85}},\ \bibinfo
  {pages} {553} (\bibinfo {year} {2013})}\BibitemShut {NoStop}%
\bibitem [{\citenamefont {Reiserer}\ and\ \citenamefont
  {Rempe}(2015)}]{2015_Rempe_RMP}%
  \BibitemOpen
  \bibfield  {author} {\bibinfo {author} {\bibfnamefont {A.}~\bibnamefont
  {Reiserer}}\ and\ \bibinfo {author} {\bibfnamefont {G.}~\bibnamefont
  {Rempe}},\ }\href {\doibase 10.1103/RevModPhys.87.1379} {\bibfield  {journal}
  {\bibinfo  {journal} {Rev. Mod. Phys.}\ }\textbf {\bibinfo {volume} {87}},\
  \bibinfo {pages} {1379} (\bibinfo {year} {2015})}\BibitemShut {NoStop}%
\bibitem [{\citenamefont {Blais}\ \emph {et~al.}(2021)\citenamefont {Blais},
  \citenamefont {Grimsmo}, \citenamefont {Girvin},\ and\ \citenamefont
  {Wallraff}}]{2021_Wallraff_RMP}%
  \BibitemOpen
  \bibfield  {author} {\bibinfo {author} {\bibfnamefont {A.}~\bibnamefont
  {Blais}}, \bibinfo {author} {\bibfnamefont {A.~L.}\ \bibnamefont {Grimsmo}},
  \bibinfo {author} {\bibfnamefont {S.~M.}\ \bibnamefont {Girvin}}, \ and\
  \bibinfo {author} {\bibfnamefont {A.}~\bibnamefont {Wallraff}},\ }\href
  {\doibase 10.1103/RevModPhys.93.025005} {\bibfield  {journal} {\bibinfo
  {journal} {Rev. Mod. Phys.}\ }\textbf {\bibinfo {volume} {93}},\ \bibinfo
  {pages} {025005} (\bibinfo {year} {2021})}\BibitemShut {NoStop}%
\bibitem [{\citenamefont {Davis}\ \emph {et~al.}(2019)\citenamefont {Davis},
  \citenamefont {Bentsen}, \citenamefont {Homeier}, \citenamefont {Li},\ and\
  \citenamefont {Schleier-Smith}}]{2019_Davis_PRL}%
  \BibitemOpen
  \bibfield  {author} {\bibinfo {author} {\bibfnamefont {E.~J.}\ \bibnamefont
  {Davis}}, \bibinfo {author} {\bibfnamefont {G.}~\bibnamefont {Bentsen}},
  \bibinfo {author} {\bibfnamefont {L.}~\bibnamefont {Homeier}}, \bibinfo
  {author} {\bibfnamefont {T.}~\bibnamefont {Li}}, \ and\ \bibinfo {author}
  {\bibfnamefont {M.~H.}\ \bibnamefont {Schleier-Smith}},\ }\href {\doibase
  10.1103/PhysRevLett.122.010405} {\bibfield  {journal} {\bibinfo  {journal}
  {Phys. Rev. Lett.}\ }\textbf {\bibinfo {volume} {122}},\ \bibinfo {pages}
  {010405} (\bibinfo {year} {2019})}\BibitemShut {NoStop}%
\bibitem [{\citenamefont {Muniz}\ \emph {et~al.}(2020)\citenamefont {Muniz},
  \citenamefont {Barberena}, \citenamefont {Lewis-Swan}, \citenamefont {Young},
  \citenamefont {Cline}, \citenamefont {Rey},\ and\ \citenamefont
  {Thompson}}]{2020_James_Nature}%
  \BibitemOpen
  \bibfield  {author} {\bibinfo {author} {\bibfnamefont {J.~A.}\ \bibnamefont
  {Muniz}}, \bibinfo {author} {\bibfnamefont {D.}~\bibnamefont {Barberena}},
  \bibinfo {author} {\bibfnamefont {R.~J.}\ \bibnamefont {Lewis-Swan}},
  \bibinfo {author} {\bibfnamefont {D.~J.}\ \bibnamefont {Young}}, \bibinfo
  {author} {\bibfnamefont {J.~R.~K.}\ \bibnamefont {Cline}}, \bibinfo {author}
  {\bibfnamefont {A.~M.}\ \bibnamefont {Rey}}, \ and\ \bibinfo {author}
  {\bibfnamefont {J.~K.}\ \bibnamefont {Thompson}},\ }\href {\doibase
  10.1038/s41586-020-2224-x} {\bibfield  {journal} {\bibinfo  {journal}
  {Nature}\ }\textbf {\bibinfo {volume} {580}},\ \bibinfo {pages} {602}
  (\bibinfo {year} {2020})}\BibitemShut {NoStop}%
\bibitem [{\citenamefont {Periwal}\ \emph {et~al.}(2021)\citenamefont
  {Periwal}, \citenamefont {Cooper}, \citenamefont {Kunkel}, \citenamefont
  {Wienand}, \citenamefont {Davis},\ and\ \citenamefont
  {Schleier-Smith}}]{2021_Monika_Nature}%
  \BibitemOpen
  \bibfield  {author} {\bibinfo {author} {\bibfnamefont {A.}~\bibnamefont
  {Periwal}}, \bibinfo {author} {\bibfnamefont {E.~S.}\ \bibnamefont {Cooper}},
  \bibinfo {author} {\bibfnamefont {P.}~\bibnamefont {Kunkel}}, \bibinfo
  {author} {\bibfnamefont {J.~F.}\ \bibnamefont {Wienand}}, \bibinfo {author}
  {\bibfnamefont {E.~J.}\ \bibnamefont {Davis}}, \ and\ \bibinfo {author}
  {\bibfnamefont {M.}~\bibnamefont {Schleier-Smith}},\ }\href {\doibase
  10.1038/s41586-021-04156-0} {\bibfield  {journal} {\bibinfo  {journal}
  {Nature}\ }\textbf {\bibinfo {volume} {600}},\ \bibinfo {pages} {630}
  (\bibinfo {year} {2021})}\BibitemShut {NoStop}%
\bibitem [{\citenamefont {Liu}\ \emph {et~al.}(2023)\citenamefont {Liu},
  \citenamefont {Wang}, \citenamefont {Yang}, \citenamefont {Wang},
  \citenamefont {Fan}, \citenamefont {Guan}, \citenamefont {Li}, \citenamefont
  {Zhang},\ and\ \citenamefont {Zhang}}]{2023_Tiancai_PRL}%
  \BibitemOpen
  \bibfield  {author} {\bibinfo {author} {\bibfnamefont {Y.}~\bibnamefont
  {Liu}}, \bibinfo {author} {\bibfnamefont {Z.}~\bibnamefont {Wang}}, \bibinfo
  {author} {\bibfnamefont {P.}~\bibnamefont {Yang}}, \bibinfo {author}
  {\bibfnamefont {Q.}~\bibnamefont {Wang}}, \bibinfo {author} {\bibfnamefont
  {Q.}~\bibnamefont {Fan}}, \bibinfo {author} {\bibfnamefont {S.}~\bibnamefont
  {Guan}}, \bibinfo {author} {\bibfnamefont {G.}~\bibnamefont {Li}}, \bibinfo
  {author} {\bibfnamefont {P.}~\bibnamefont {Zhang}}, \ and\ \bibinfo {author}
  {\bibfnamefont {T.}~\bibnamefont {Zhang}},\ }\href {\doibase
  10.1103/PhysRevLett.130.173601} {\bibfield  {journal} {\bibinfo  {journal}
  {Phys. Rev. Lett.}\ }\textbf {\bibinfo {volume} {130}},\ \bibinfo {pages}
  {173601} (\bibinfo {year} {2023})}\BibitemShut {NoStop}%
\bibitem [{\citenamefont {Coslovich}\ \emph {et~al.}(2006)\citenamefont
  {Coslovich}, \citenamefont {Pesenti},\ and\ \citenamefont
  {Ukovich}}]{2006_Ukovich_NPPapplication}%
  \BibitemOpen
  \bibfield  {author} {\bibinfo {author} {\bibfnamefont {L.}~\bibnamefont
  {Coslovich}}, \bibinfo {author} {\bibfnamefont {R.}~\bibnamefont {Pesenti}},
  \ and\ \bibinfo {author} {\bibfnamefont {W.}~\bibnamefont {Ukovich}},\ }\href
  {\doibase 10.1080/13928619.2006.9637717} {\bibfield  {journal} {\bibinfo
  {journal} {Ukio Technologinis ir Ekonominis Vystymas}\ }\textbf {\bibinfo
  {volume} {12}},\ \bibinfo {pages} {18} (\bibinfo {year} {2006})}\BibitemShut
  {NoStop}%
\bibitem [{\citenamefont {Perdomo-Ortiz}\ \emph {et~al.}(2011)\citenamefont
  {Perdomo-Ortiz}, \citenamefont {Venegas-Andraca},\ and\ \citenamefont
  {Aspuru-Guzik}}]{2011_Aspuru_RQC}%
  \BibitemOpen
  \bibfield  {author} {\bibinfo {author} {\bibfnamefont {A.}~\bibnamefont
  {Perdomo-Ortiz}}, \bibinfo {author} {\bibfnamefont {S.~E.}\ \bibnamefont
  {Venegas-Andraca}}, \ and\ \bibinfo {author} {\bibfnamefont {A.}~\bibnamefont
  {Aspuru-Guzik}},\ }\href {\doibase 10.1007/s11128-010-0168-z} {\bibfield
  {journal} {\bibinfo  {journal} {Quantum Information Processing}\ }\textbf
  {\bibinfo {volume} {10}},\ \bibinfo {pages} {33} (\bibinfo {year}
  {2011})}\BibitemShut {NoStop}%
\bibitem [{\citenamefont {Chancellor}(2017)}]{2017_Chancellor_NJP}%
  \BibitemOpen
  \bibfield  {author} {\bibinfo {author} {\bibfnamefont {N.}~\bibnamefont
  {Chancellor}},\ }\href {\doibase 10.1088/1367-2630/aa59c4} {\bibfield
  {journal} {\bibinfo  {journal} {New Journal of Physics}\ }\textbf {\bibinfo
  {volume} {19}},\ \bibinfo {pages} {023024} (\bibinfo {year}
  {2017})}\BibitemShut {NoStop}%
\bibitem [{\citenamefont {{Yamashiro}}\ \emph {et~al.}(2019)\citenamefont
  {{Yamashiro}}, \citenamefont {{Ohkuwa}}, \citenamefont {{Nishimori}},\ and\
  \citenamefont {{Lidar}}}]{2019_Lidar_RQA}%
  \BibitemOpen
  \bibfield  {author} {\bibinfo {author} {\bibfnamefont {Y.}~\bibnamefont
  {{Yamashiro}}}, \bibinfo {author} {\bibfnamefont {M.}~\bibnamefont
  {{Ohkuwa}}}, \bibinfo {author} {\bibfnamefont {H.}~\bibnamefont
  {{Nishimori}}}, \ and\ \bibinfo {author} {\bibfnamefont {D.~A.}\ \bibnamefont
  {{Lidar}}},\ }\href {\doibase 10.1103/PhysRevA.100.052321} {\bibfield
  {journal} {\bibinfo  {journal} {\pra}\ }\textbf {\bibinfo {volume} {100}},\
  \bibinfo {eid} {052321} (\bibinfo {year} {2019})}\BibitemShut {NoStop}%
\bibitem [{\citenamefont {Gra{\ss}}\ \emph {et~al.}(2016)\citenamefont
  {Gra{\ss}}, \citenamefont {Ravent{\'o}s}, \citenamefont
  {Juli{\'a}-D{\'\i}az}, \citenamefont {Gogolin},\ and\ \citenamefont
  {Lewenstein}}]{2016_Lewenstein_NC}%
  \BibitemOpen
  \bibfield  {author} {\bibinfo {author} {\bibfnamefont {T.}~\bibnamefont
  {Gra{\ss}}}, \bibinfo {author} {\bibfnamefont {D.}~\bibnamefont
  {Ravent{\'o}s}}, \bibinfo {author} {\bibfnamefont {B.}~\bibnamefont
  {Juli{\'a}-D{\'\i}az}}, \bibinfo {author} {\bibfnamefont {C.}~\bibnamefont
  {Gogolin}}, \ and\ \bibinfo {author} {\bibfnamefont {M.}~\bibnamefont
  {Lewenstein}},\ }\href {\doibase 10.1038/ncomms11524} {\bibfield  {journal}
  {\bibinfo  {journal} {Nature Communications}\ }\textbf {\bibinfo {volume}
  {7}},\ \bibinfo {pages} {11524} (\bibinfo {year} {2016})}\BibitemShut
  {NoStop}%
\bibitem [{\citenamefont {Anikeeva}\ \emph {et~al.}(2021)\citenamefont
  {Anikeeva}, \citenamefont {Markovi{\'c}}, \citenamefont {Borish},
  \citenamefont {Hines}, \citenamefont {Rajagopal}, \citenamefont {Cooper},
  \citenamefont {Periwal}, \citenamefont {Safavi-Naeini}, \citenamefont
  {Davis},\ and\ \citenamefont {Schleier-Smith}}]{anikeeva2021number}%
  \BibitemOpen
  \bibfield  {author} {\bibinfo {author} {\bibfnamefont {G.}~\bibnamefont
  {Anikeeva}}, \bibinfo {author} {\bibfnamefont {O.}~\bibnamefont
  {Markovi{\'c}}}, \bibinfo {author} {\bibfnamefont {V.}~\bibnamefont
  {Borish}}, \bibinfo {author} {\bibfnamefont {J.~A.}\ \bibnamefont {Hines}},
  \bibinfo {author} {\bibfnamefont {S.~V.}\ \bibnamefont {Rajagopal}}, \bibinfo
  {author} {\bibfnamefont {E.~S.}\ \bibnamefont {Cooper}}, \bibinfo {author}
  {\bibfnamefont {A.}~\bibnamefont {Periwal}}, \bibinfo {author} {\bibfnamefont
  {A.}~\bibnamefont {Safavi-Naeini}}, \bibinfo {author} {\bibfnamefont {E.~J.}\
  \bibnamefont {Davis}}, \ and\ \bibinfo {author} {\bibfnamefont
  {M.}~\bibnamefont {Schleier-Smith}},\ }\href@noop {} {\bibfield  {journal}
  {\bibinfo  {journal} {PRX Quantum}\ }\textbf {\bibinfo {volume} {2}},\
  \bibinfo {pages} {020319} (\bibinfo {year} {2021})}\BibitemShut {NoStop}%
\bibitem [{\citenamefont {A.~Yu~Kitaev}(2000)}]{Kitaev_QCBook}%
  \BibitemOpen
  \bibfield  {author} {\bibinfo {author} {\bibfnamefont {M.~V.}\ \bibnamefont
  {A.~Yu~Kitaev}, \bibfnamefont {A.~H.~Shen}},\ }\href@noop {} {\emph {\bibinfo
  {title} {Classical and Quantum Computation}}},\ Vol.~\bibinfo {volume} {47}\
  (\bibinfo  {publisher} {American Mathematical Society},\ \bibinfo {year}
  {2000})\BibitemShut {NoStop}%
\bibitem [{\citenamefont {Shah}\ \emph {et~al.}(2007)\citenamefont {Shah},
  \citenamefont {LAM},\ and\ \citenamefont {ROWLAND}}]{2007_lam_mathematics}%
  \BibitemOpen
  \bibfield  {author} {\bibinfo {author} {\bibfnamefont {S.}~\bibnamefont
  {Shah}}, \bibinfo {author} {\bibfnamefont {T.}~\bibnamefont {LAM}}, \ and\
  \bibinfo {author} {\bibfnamefont {D.}~\bibnamefont {ROWLAND}},\ }\href@noop
  {} {\bibfield  {journal} {\bibinfo  {journal} {Mathematics Review}\ }
  (\bibinfo {year} {2007})}\BibitemShut {NoStop}%
\bibitem [{\citenamefont {Cormen}\ \emph {et~al.}(2022)\citenamefont {Cormen},
  \citenamefont {Leiserson}, \citenamefont {Rivest},\ and\ \citenamefont
  {Stein}}]{cormen2022introduction}%
  \BibitemOpen
  \bibfield  {author} {\bibinfo {author} {\bibfnamefont {T.~H.}\ \bibnamefont
  {Cormen}}, \bibinfo {author} {\bibfnamefont {C.~E.}\ \bibnamefont
  {Leiserson}}, \bibinfo {author} {\bibfnamefont {R.~L.}\ \bibnamefont
  {Rivest}}, \ and\ \bibinfo {author} {\bibfnamefont {C.}~\bibnamefont
  {Stein}},\ }\href@noop {} {\emph {\bibinfo {title} {Introduction to
  algorithms}}}\ (\bibinfo  {publisher} {MIT press},\ \bibinfo {year}
  {2022})\BibitemShut {NoStop}%
\bibitem [{\citenamefont {Kochenberger}\ \emph {et~al.}(2014)\citenamefont
  {Kochenberger}, \citenamefont {Hao}, \citenamefont {Glover}, \citenamefont
  {Lewis}, \citenamefont {L{\"u}}, \citenamefont {Wang},\ and\ \citenamefont
  {Wang}}]{2014_Wang_QUBO}%
  \BibitemOpen
  \bibfield  {author} {\bibinfo {author} {\bibfnamefont {G.}~\bibnamefont
  {Kochenberger}}, \bibinfo {author} {\bibfnamefont {J.-K.}\ \bibnamefont
  {Hao}}, \bibinfo {author} {\bibfnamefont {F.}~\bibnamefont {Glover}},
  \bibinfo {author} {\bibfnamefont {M.}~\bibnamefont {Lewis}}, \bibinfo
  {author} {\bibfnamefont {Z.}~\bibnamefont {L{\"u}}}, \bibinfo {author}
  {\bibfnamefont {H.}~\bibnamefont {Wang}}, \ and\ \bibinfo {author}
  {\bibfnamefont {Y.}~\bibnamefont {Wang}},\ }\href@noop {} {\bibfield
  {journal} {\bibinfo  {journal} {Journal of combinatorial optimization}\
  }\textbf {\bibinfo {volume} {28}},\ \bibinfo {pages} {58} (\bibinfo {year}
  {2014})}\BibitemShut {NoStop}%
\bibitem [{\citenamefont {Santoro}\ \emph {et~al.}(2002)\citenamefont
  {Santoro}, \citenamefont {Martoňák}, \citenamefont {Tosatti},\ and\
  \citenamefont {Car}}]{2002_Roberto_Science}%
  \BibitemOpen
  \bibfield  {author} {\bibinfo {author} {\bibfnamefont {G.~E.}\ \bibnamefont
  {Santoro}}, \bibinfo {author} {\bibfnamefont {R.}~\bibnamefont {Martoňák}},
  \bibinfo {author} {\bibfnamefont {E.}~\bibnamefont {Tosatti}}, \ and\
  \bibinfo {author} {\bibfnamefont {R.}~\bibnamefont {Car}},\ }\href {\doibase
  10.1126/science.1068774} {\bibfield  {journal} {\bibinfo  {journal}
  {Science}\ }\textbf {\bibinfo {volume} {295}},\ \bibinfo {pages} {2427}
  (\bibinfo {year} {2002})}\BibitemShut {NoStop}%
\bibitem [{\citenamefont {Lechner}\ \emph {et~al.}(2015)\citenamefont
  {Lechner}, \citenamefont {Hauke},\ and\ \citenamefont
  {Zoller}}]{2015_Zoller_SciAdv}%
  \BibitemOpen
  \bibfield  {author} {\bibinfo {author} {\bibfnamefont {W.}~\bibnamefont
  {Lechner}}, \bibinfo {author} {\bibfnamefont {P.}~\bibnamefont {Hauke}}, \
  and\ \bibinfo {author} {\bibfnamefont {P.}~\bibnamefont {Zoller}},\ }\href
  {\doibase 10.1126/sciadv.1500838} {\bibfield  {journal} {\bibinfo  {journal}
  {Science Advances}\ }\textbf {\bibinfo {volume} {1}},\ \bibinfo {pages}
  {e1500838} (\bibinfo {year} {2015})}\BibitemShut {NoStop}%
\bibitem [{\citenamefont {Pastawski}\ and\ \citenamefont
  {Preskill}(2016)}]{2016_Preskill_ParityDecoder}%
  \BibitemOpen
  \bibfield  {author} {\bibinfo {author} {\bibfnamefont {F.}~\bibnamefont
  {Pastawski}}\ and\ \bibinfo {author} {\bibfnamefont {J.}~\bibnamefont
  {Preskill}},\ }\href {\doibase 10.1103/PhysRevA.93.052325} {\bibfield
  {journal} {\bibinfo  {journal} {Phys. Rev. A}\ }\textbf {\bibinfo {volume}
  {93}},\ \bibinfo {pages} {052325} (\bibinfo {year} {2016})}\BibitemShut
  {NoStop}%
\bibitem [{\citenamefont {Hauke}\ \emph {et~al.}(2020)\citenamefont {Hauke},
  \citenamefont {Katzgraber}, \citenamefont {Lechner}, \citenamefont
  {Nishimori},\ and\ \citenamefont {Oliver}}]{hauke2020perspectives}%
  \BibitemOpen
  \bibfield  {author} {\bibinfo {author} {\bibfnamefont {P.}~\bibnamefont
  {Hauke}}, \bibinfo {author} {\bibfnamefont {H.~G.}\ \bibnamefont
  {Katzgraber}}, \bibinfo {author} {\bibfnamefont {W.}~\bibnamefont {Lechner}},
  \bibinfo {author} {\bibfnamefont {H.}~\bibnamefont {Nishimori}}, \ and\
  \bibinfo {author} {\bibfnamefont {W.~D.}\ \bibnamefont {Oliver}},\
  }\href@noop {} {\bibfield  {journal} {\bibinfo  {journal} {Reports on
  Progress in Physics}\ }\textbf {\bibinfo {volume} {83}},\ \bibinfo {pages}
  {054401} (\bibinfo {year} {2020})}\BibitemShut {NoStop}%
\bibitem [{\citenamefont {Tanji-Suzuki}\ \emph {et~al.}()\citenamefont
  {Tanji-Suzuki}, \citenamefont {Leroux}, \citenamefont {Schleier-Smith},
  \citenamefont {Cetina}, \citenamefont {Grier}, \citenamefont {Simon},\ and\
  \citenamefont {Vuletić}}]{arimondo_chapter_2011}%
  \BibitemOpen
  \bibfield  {author} {\bibinfo {author} {\bibfnamefont {H.}~\bibnamefont
  {Tanji-Suzuki}}, \bibinfo {author} {\bibfnamefont {I.~D.}\ \bibnamefont
  {Leroux}}, \bibinfo {author} {\bibfnamefont {M.~H.}\ \bibnamefont
  {Schleier-Smith}}, \bibinfo {author} {\bibfnamefont {M.}~\bibnamefont
  {Cetina}}, \bibinfo {author} {\bibfnamefont {A.~T.}\ \bibnamefont {Grier}},
  \bibinfo {author} {\bibfnamefont {J.}~\bibnamefont {Simon}}, \ and\ \bibinfo
  {author} {\bibfnamefont {V.}~\bibnamefont {Vuletić}},\ }in\ \href {\doibase
  https://doi.org/10.1016/B978-0-12-385508-4.00004-8} {\emph {\bibinfo
  {booktitle} {Advances in Atomic, Molecular, and Optical Physics}}},\
  Vol.~\bibinfo {volume} {60},\ \bibinfo {editor} {edited by\ \bibinfo {editor}
  {\bibfnamefont {E.}~\bibnamefont {Arimondo}}, \bibinfo {editor}
  {\bibfnamefont {P.~R.}\ \bibnamefont {Berman}}, \ and\ \bibinfo {editor}
  {\bibfnamefont {C.~C.}\ \bibnamefont {Lin}}}\ (\bibinfo  {publisher}
  {Academic Press})\ pp.\ \bibinfo {pages} {201--237},\ \bibinfo {note}
  {{ISSN}: 1049-250X}\BibitemShut {NoStop}%
\end{thebibliography}%
\bibliographystyle{apsrev4-1}

\newpage

\begin{widetext} 
\renewcommand{\theequation}{S\arabic{equation}}
\renewcommand{\thesection}{S-\arabic{section}}
\renewcommand{\thefigure}{S\arabic{figure}}
\renewcommand{\thetable}{S\arabic{table}}
\setcounter{equation}{0}
\setcounter{figure}{0}
\setcounter{table}{0}

\newpage 

\begin{center}
    \Huge Supplementary Material
\end{center}

\section{Effective atomic  couplings by the  Raman scheme in the optical cavity}
We derive the effective atomic couplings introduced in the main text in this section. The original experimental Hamiltonian is 
\begin{align}
  \begin{split}
    \hat{H}_{\rm exp}/\hbar =& \omega \hat{a}^\dagger \hat{a} + \sum_i \left\{ -\omega_0 \ket{\down}_i\!\bra{\down} + (-\omega_0+\Delta_F)\ket{\up}_i\!\bra{\up} \right\} \\
    &+ \sum_i \left\{ \Omega_1 \cos(\omega_1 t) \ket{e}_i\!\bra{\down} + \Omega_2 \cos(\omega_2 t) \ket{e}_i\!\bra{\up} +\hc \right\} \\
    &+ \sum_i \lambda_i g_0 \left\{ \hat{a} \ket{e}_i\!\bra{\down} + \hat{a} \ket{e}_i\!\bra{\up} +\hc \right\} \\
  \end{split},
\end{align}
where the frequency definitions are given in Fig.~\ref{fig:experiment} in the main text, and the energy zero point is set to $E(\ket{e})=0$. Here we neglect the effects of $\Omega_{1(2)}$ driving the transition $\ket{\up} (\ket{\down}) \rightarrow \ket{e}$. We also assume that the atoms are tightly trapped along the propagation direction of the coupling beams $\Omega_{1,2}$, so that the phases of the coupling beams are identical across the entire atomic ensemble.
We decompose the Hamiltonian into $\hat{H}_{\rm exp} = H_0 + V_1 + V_2$, where
\begin{align}
  \hat{H}_0/\hbar &\equiv \omega \hat{a}^\dagger \hat{a} + \sum_i \left\{ -\omega_0 \ket{\down}_i\!\bra{\down} + (-\omega_0+\Delta_F)\ket{\up}_i\!\bra{\up} \right\}, \\
  \hat{V}_1/\hbar &\equiv \sum_i \left\{ \Omega_1 \cos(\omega_1 t) \ket{e}_i\!\bra{\down} + \lambda_i g_0 \hat{a} \ket{e}_i\!\bra{\up} +\hc \right\}, \\
  \hat{V}_2/\hbar &\equiv \sum_i \left\{ \Omega_2 \cos(\omega_2 t) \ket{e}_i\!\bra{\up} + \lambda_i g_0 \hat{a} \ket{e}_i\!\bra{\down} +\hc \right\}.
\end{align} 

{In the following we assume, $|\Omega_1|\sim |\Omega_2|\sim | \lambda g_0 | \ll |\Delta|\sim |\Delta_F|$, and treat $\hat{V}_1$ and $\hat{V}_2$ as perturbations.}  
We first perform a rotating wave transformation through  $\hat{U}_1 = \exp(i \hat{T}_{\rm rot,1}t/\hbar)$, where
\begin{align}
\hat{T}_{\rm rot,1}/\hbar = \omega \hat{a}^\dagger\hat{a} - \sum_i \left\{ \omega_1 \ket{\down}_i\!\bra{\down} + \omega \ket{\up}_i\!\bra{\up} \right\}.
\end{align}
This results in the Hamiltonian in the rotating frame, after rotating wave approximation (RWA), 
\begin{align}
  \begin{split}
    \hat{H}_{\rm rot, 1} /\hbar &= \hat{U}_1 \hat{H}_{\rm exp} \hat{U}_1^\dagger + i\left( \partial_t \hat{U}_1 \right) \hat{U}_1^\dagger \\
    &= \sum_i \left\{ (\omega_1-\omega_0) \ket{\down}_i\!\bra{\down} + (\omega-\omega_0+\Delta_F) \ket{\up}_i\!\bra{\up} \right\} \\
    &\qquad + \sum_i \left\{ \frac{\Omega_1}{2} \ket{e}_i\!\bra{\down} + \lambda_i g_0 \hat{a} \ket{e}_i\!\bra{\up} + \hc \right\} \\
    &\qquad + \hat{U}_1 \hat{V}_2 \hat{U}_1^\dagger
  \end{split}.
  \label{eq:exp H rot1}
\end{align}
{We then perform a Van-Vleck transformation
 $\hat{H} _{{\rm rot}, 1}' =   e^{-i\hat{S}_1} \hat{H}_{\rm rot,1} e^{i\hat{S}_1}$, with 
 \be 
\hat{S}_1 = \sum_i \left[ 
	 \frac{i\Omega_1 /2}{\omega_1-\omega_0} |\downarrow\rangle_i \langle e| 
	+\frac{i\lambda_ig_0}{\omega-\omega_0+\Delta_F}\hat{ a}^\dag |\uparrow\rangle_i \langle e| + h.c. \right].    
 \ee 
 } 
It follows that 
\bea 
\hat{H}_{\rm rot, 1} '/\hbar
&=& \sum_i \left\{ (\omega_1-\omega_0) \ket{\down}_i\!\bra{\down} + (\omega-\omega_0+\Delta_F) \ket{\up}_i\!\bra{\up} \right\} \\
&+& \sum_i \left[  \left( \frac{\Omega_1^2/4}{\omega_1-\omega_0}\right) \ket{\downarrow}_i \bra{ \downarrow} 
+ \left( \frac{\lambda_i ^2 g_0^2}{\omega - \omega_0+ \Delta_F} \right) \hat{a} ^\dagger \ket{ \uparrow} _i \bra{\uparrow} \hat{a}  
\right]   \nn \\
&+& \frac{1}{4} \sum_i \left[ 
		  \frac{\Omega_1 \lambda_i g_0 (\omega + \omega_1 -2\omega_0 + \Delta_F) }{(\omega_1-\omega_0) (\omega-\omega_0+\Delta_F) }  
			 \hat{a} ^\dagger \ket{ \uparrow} \bra{\downarrow}   + h.c. \right]   \nn  \\ 
&+& e^{-i\hat{S}_1} \hat{U}_1 \hat{V}_2 \hat{U}_1^\dagger e^{i\hat{S}_1} 
\eea
With the frequencies   
$\omega_1 = \omega_0 + \Delta + \Delta_F - \delta - \delta_m$, 
$\omega_2 = \omega_0 + \Delta - \Delta_F - \delta$, 
$ \omega = \omega_0 + \Delta$, and $|\delta|$, $|\delta_m| \ll |\Delta|, |\Delta_F|$, 
the crossing terms between $\hat{V}_1$ and $\hat{V}_2$ are all far off-resonant. 
We then have $e^{-i\hat{S}_1} \hat{U}_1 \hat{V}_2 \hat{U}_1^\dagger e^{i\hat{S}_1} \approx \hat{U}_1 \hat{V}_2 \hat{U}_1^\dagger$

We now perform a second rotating wave transformation through  $\hat{U}_2 = \exp(i \hat{T}_{\rm rot, 2}t/\hbar)$, 
with 
\be 
  \hat{T}_{\rm rot, 2}/\hbar= - \sum_i \left\{ (\omega-\omega_1) \ket{\down}_i\!\bra{\down} +(\omega_2-\omega) \ket{\up}_i\!\bra{\up} \right\}. 
\ee 
The transformed Hamiltonian reads as 
\bea
\hat{H}_{\rm rot, 2} /\hbar
&=&  \sum_i \left\{ (\omega -\omega_0) \ket{\down}_i\!\bra{\down} + (\omega_2 -\omega_0+\Delta_F) \ket{\up}_i\!\bra{\up} \right\} 
\\
&+& \sum_i \left[  \left( \frac{\Omega_1^2/4}{\omega_1-\omega_0}\right) \ket{\downarrow}_i \bra{ \downarrow} 
+ \left( \frac{\lambda_i ^2 g_0^2}{\omega - \omega_0+ \Delta_F} \right) \hat{a} ^\dagger \ket{ \uparrow} _i \bra{\uparrow} \hat{a}  
\right]   \nn \\
&+& \frac{1}{4} \sum_i \left[ 
		  \frac{\Omega_1 \lambda_i g_0 (\omega + \omega_1 -2\omega_0 + \Delta_F) }{(\omega_1-\omega_0) (\omega-\omega_0+\Delta_F) }  
			 e^{-i(\omega_1+ \omega_2 - 2\omega)t}  
			 \hat{a} ^\dagger \ket{ \uparrow} \bra{\downarrow}   + h.c. \right]   \nn  \\ 
&+&  \sum_i \left\{ \frac{\Omega_2}{2}  \ket{e}_i\!\bra{\up} + \lambda_i g_0 \hat{a} \ket{e}_i\!\bra{\down} +\hc \right\}.
\eea 
Introducing a second  Van-Vleck transformation
 $\hat{H} _{{\rm rot}, 2}' =   e^{-i\hat{S}_2} \hat{H}_{\rm rot,2  } e^{i\hat{S}_2}$, with 
 \be 
\hat{S} _2  = 
\sum_i \left[ 
	 \frac{i\Omega_2 /2}{\omega_2-\omega_0+\Delta_F} \ket{\up} _i \langle e| 
	+\frac{i\lambda_ig_0}{\omega-\omega_0} \hat{a}^\dag \ket{\down}_i \langle e| + h.c. \right], 
 \ee 
we obtain 
\bea 
\hat{H}_{\rm rot, 2} '/\hbar
&=& \sum_i \left\{ (\omega -\omega_0) \ket{\down}_i\!\bra{\down} + (\omega_2-\omega_0+\Delta_F) \ket{\up}_i\!\bra{\up} \right\} \\
&+& \sum_i \left[  \left( \frac{\Omega_1^2/4}{\omega_1-\omega_0}\right) \ket{\downarrow}_i \bra{ \downarrow} 
+ \left( \frac{\lambda_i ^2 g_0^2}{\omega - \omega_0+ \Delta_F} \right) \hat{a}^\dagger \ket{ \uparrow} _i \bra{\uparrow} \hat{a}  
\right]   \nn \\
&+&\sum_i \left[  \left( \frac{\Omega_2^2/4}{\omega_2-\omega_0 + \Delta_F }\right) \ket{\up}_i \bra{ \up} 
+ \left( \frac{\lambda_i^2 g_0^2}{\omega-\omega_0} \right)  \hat{a}^\dagger \ket{\down} _i \bra{\down} \hat{a}  
\right]   \nn \\
&+& \frac{1}{4} \sum_i \left[ 
		  \frac{\Omega_1 \lambda_i g_0 (\omega + \omega_1 -2\omega_0 + \Delta_F) }{(\omega_1-\omega_0) (\omega-\omega_0+\Delta_F) }  
		  e^{-i(\omega_1+ \omega_2 - 2\omega)t}  
			 \hat{a} ^\dagger \ket{ \uparrow} \bra{\downarrow}   + h.c. \right]   \nn  \\ 
&+&   \frac{1}{4} \sum_i \left[ 
		  \frac{\Omega_2 \lambda_i g_0 (\omega + \omega_2 -2\omega_0 + \Delta_F) }{(\omega -\omega_0) (\omega_2-\omega_0+\Delta_F) }  
			 \hat{a}^\dagger \ket{ \down} \bra{\up}   + h.c. \right]  
\eea 
In a proper rotating wave frame, the effective Hamiltonian takes a form 
\bea
&& H_{\rm exp}' /\hbar \nn \\
&\approx &
\delta  \hat{a}^\dag \hat{a} \nn \\ 
&+&   \sum_i \left\{ - \frac{\delta_m}{2} \ket{\down}_i\!\bra{\down}  
+ \frac{1}{4}  \left[  \left( \frac{\Omega_1^2}{\Delta + \Delta_F }\right) \ket{\downarrow}_i\! \bra{ \downarrow}   
+  \left( \frac{\Omega_2^2 }{\Delta }\right)\! \ket{\up}_i \bra{ \up} 
\right]  \right\}   \nn \\
&+& \frac{1}{2} \sum_i \left\{ 
		  \left( \frac{\Omega_1 \lambda_i g_0  }{\Delta+\Delta_F}  \right) 
			 \hat{a}^\dagger \ket{ \uparrow}_i \!\bra{\downarrow}   + h.c. \right\}   \nn  \\ 
&+&   \frac{1}{2} \sum_i \left\{  
		\left(  \frac{\Omega_2 \lambda_i g_0  }{\Delta }  \right) 
			 \hat{a}^\dagger \ket{ \down}_i \!\bra{\up}   + h.c. \right\}, 
\eea 
 where we have assumed the cavity photon number is negligible. This could be guaranteed by choosing $|\delta_m|\sim |{\Omega_{1,2} \lambda_i g_0}/{\Delta} |  \ll |\delta| \ll |\Delta|, |\Delta_F|$. 
{In the atom cavity system, it is reasonable to set $\Delta\sim$GHz, $\Delta_F =6.8$ GHz (for $^{87}$Rb atoms), 
and $\Omega_{1,2} \sim \delta\sim 10^2$ MHz, and $g_0\sim10$ MHz.}  
The resultant energy splitting between $\ket{\up}$ and $\ket{\down}$ is 
\be 
2 \tilde{\delta} _m = \frac{\delta_m}{2} 
+\frac{\Omega_2^2 }{4\Delta }
- \frac{\Omega_1^2}{4(\Delta + \Delta_F) } 
\ee

With $g_{R,1}\equiv \Omega_1g_0/2(\Delta+\Delta_F)$, $g_{R,2}\equiv\Omega_2g_0/2\Delta$, 
the Hamiltonian above can be written as 
\begin{align}
  \begin{split}
    \hat{H}_{\rm exp}^{\prime} /\hbar
    &\approx \delta\hat{a}^\dagger\hat{a} + \tilde{\delta}_m \sum_i \hat{\sigma}_z^i \\
      &\qquad+ \sum_i \left\{  \lambda_i g_{R,1} \hat{a}\ket{\down}_i\!\bra{\up} + \lambda_i g_{R,2} \hat{a}^\dagger  \ket{\down}_i\!\bra{\up} + \hc \right\}.
  \end{split}
\end{align}
We assume $|g_{R, 1(2)}| \ll |\delta|$, and perform another Van-Vleck transformation to decouple the cavity mode, 
$e^{-i\hat{S}_3}  \hat{H}_{\rm exp}' e^{i\hat{S}_3}$, 
with 
\be
\hat{S}_3 =\frac{1}{\delta}  \sum_j \left[ i\lambda_j  g_{R,1} \hat{a}^\dag\ket{\up}_j \!\bra{\down} 
				   -i \lambda_j g_{R,2} \hat{a} \ket{\up}_j\! \ket{\down}+h.c. \right]. 
\ee 
With $\Omega_1/\Omega_2=(\Delta+\Delta_F)/\Delta$, we have $g_{R,1}=g_{R,2}\equiv g_{R}$. This leads to a fourth-order effective Hamiltonian
\begin{align}
  \begin{split}
  H_{\rm eff}=\tilde{\delta}_m \sum_i \hat{\sigma}^z_i + g_4 \left(\sum_i{\lambda_i}\hat{\sigma}^x_i\right)^2,
  \end{split}
\end{align}
where $g_4=-g_R^2/\delta$.


\section{Scaling analysis of the quantum coherence time}

In the four-photon Raman scheme, the decoherence is introduced by the finite lifetime of the atomic excited state and the cavity photon leakage. Here, we give a scaling analysis and show that the quantum coherence time is determined by the cavity cooperativity. First, both of the driving beam and the cavity photons induce AC-Stark shift to the ground states $\ket{\up}$ and $\ket{\down}$. Due to the spontaneous emission, the shifted ground states have effective lifetime according to the imaginary part of the AC-Stark shift. 
In our proposal, we assume cavity photons are virtually excited to mediate atomic interactions, 
so the photon number is negligible, i.e., $\langle \hat{a}^\dagger\hat{a} \rangle \approx 0$. 
The ground state linewidth induced by the Raman coupling is then approximately given by 
\begin{align}
\gamma_1 \sim \frac{\Omega^2}{\Delta^2}\frac{\Gamma}{2},
\end{align}
where $\Omega \sim \Omega_{1,2}$, $\Delta$, $\Gamma$, represent the Rabi frequencies of the Raman coupling, the single-photon detunings, and the linewidth of the excited state, respectively. Second, the spontaneous emission also generates a non-zero imaginary part of the two-photon coupling Rabi frequency $\sim g_R (\Gamma/2\Delta)$, and leads to an  imaginary part of the four-photon Rabi frequency $\gamma_2 \sim g_4 (\Gamma/\Delta)$. Third, the cavity photon leakage generates a non-zero imaginary part of the four-photon Rabi frequency $\gamma_3 \sim g_4 (\kappa/2\delta)$, where $\kappa$ is the cavity linewidth. The worst-scenario estimate  corresponds to  putting the system always in its brightest state, leading to an upper-bound of the decoherence rate given by the sum of the three terms, $\gamma=\gamma_1+\gamma_2+\gamma_3$.

With an approximate expression for the four-photon Rabi frequency $g_4 \sim (\Omega g_0 / 2\Delta)^2/\delta$, the relative decoherence rates of the three terms are as follows. 
\begin{align}
  \frac{\gamma_1}{g_4} \sim \frac{2\delta\Gamma}{g_0^2} = \frac{8}{\eta} \frac{\delta}{\kappa},
\end{align}
where $\eta = 4g_0^2/\Gamma\kappa$ is the cavity cooperativity.
The other two decoherence rates are 
\begin{align}
  \frac{\gamma_2}{g_4} \sim \frac{\Gamma}{\Delta},
\end{align}
and
\begin{align}
  \frac{\gamma_3}{g_4} \sim \frac{\kappa}{2\delta}.
\end{align}
It is now clear that the $\gamma_2$ can be {efficiently suppressed} by large single-photon detunings. At the same time, there is a trade off between $\gamma_1$ and $\gamma_3$. The best performance is achieved when $\gamma_1 = \gamma_3$, i.e., $\delta/\kappa = \sqrt{\eta}/4$, which is reasonable in experiments with $\eta \gtrsim 100$. Then we have the quantum coherence time, up to a coeffcicient of order unity,
\begin{align}
  N g_4 T \sim \frac{1}{\gamma} \sim \frac{1}{4}\sqrt{\eta}.
\end{align}

\begin{figure*}[htp]
  \begin{center}
  \includegraphics[width=.8\linewidth]{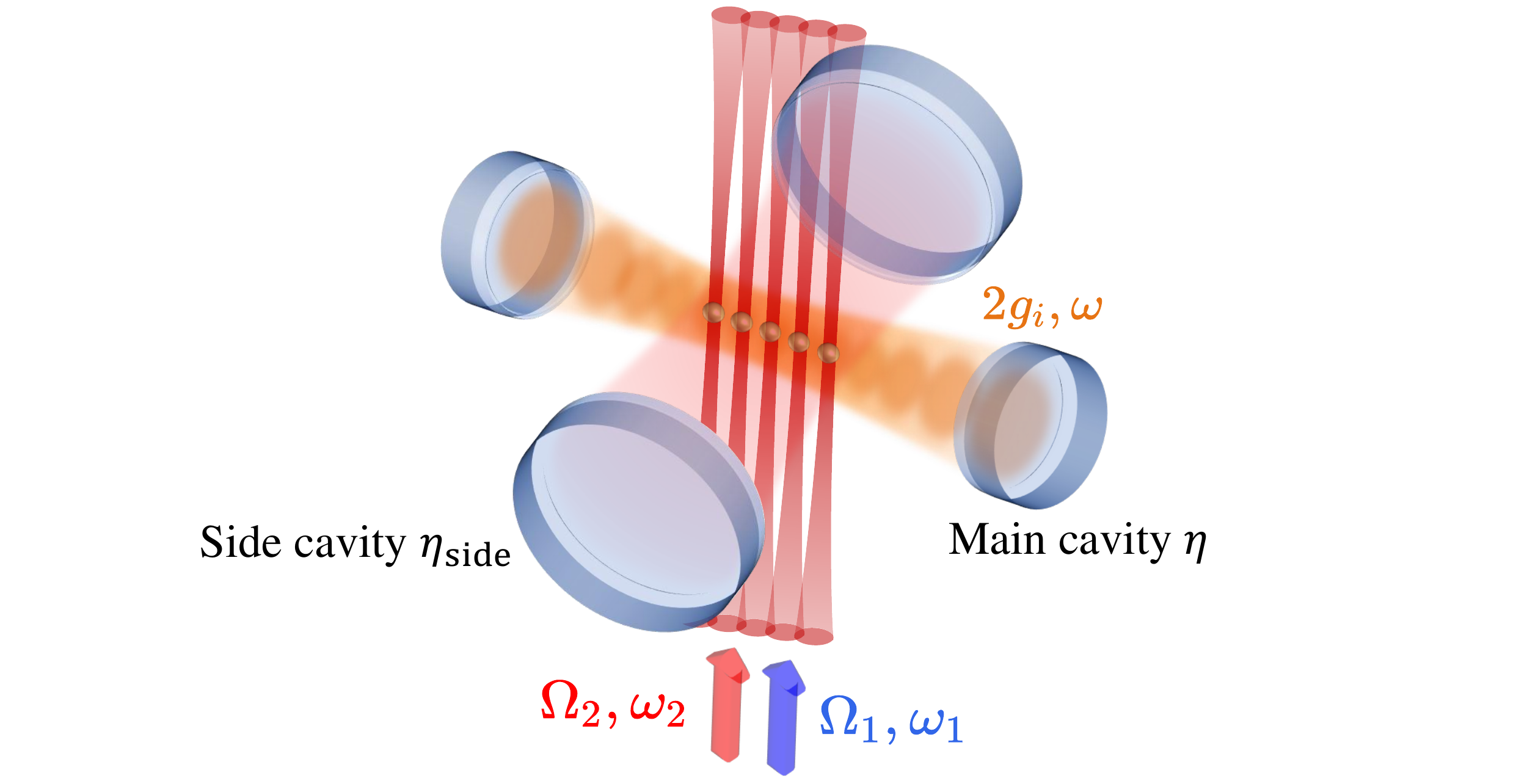}
  \end{center}
  \caption{\label{fig:side-cavity} Side cavity setup for scaling up. Compared to the scheme described in the main text, the atoms are placed in an additional side cavity with cooperativity $\eta_{\rm side}$. The coupling beams are sent along with the tweezer beams for local addressing. The photons scattered by the atoms are captured by the side cavity and participate in the coherent interaction again, leading to an effective boost to the cooperativity of the main cavity. The atoms are placed at the anti-nodes of the side cavity to achieve the best performance allowed by the collective radiation of the atoms into the cavity mode.
  }
\end{figure*}

The quantum coherence time in the scheme above scales with atom number as $1/N$. This deteriorates when $N$ becomes larger. 
{We thus provide an alternative scheme to improve the scalability of our scheme,} 
by putting the atoms at the anti-nodes of a side-cavity, as shown in Fig.~\ref{fig:side-cavity}. The coupling beams $\Omega_{1,2}$ are sent in along with the tweezer beams for local addressing. We follow Ref.~\cite{arimondo_chapter_2011} and model the atomic spontaneous emission as the coupling beam photons being scattered by the atoms. With all of the atoms placed at the anti-nodes of the side cavity, 
the scattering is suppressed by a factor of
\begin{align}
  \frac{1}{ \left[ 1 + N\eta_{\rm side}\mathcal{L}_a(\Delta/\Gamma) \right]^2 +  \left[ N\eta_{\rm side}\mathcal{L}_d(\Delta/\Gamma) \right]^2},
\end{align}
where $\eta_{\rm side}$ is the cooperativity of the side cavity, $\mathcal{L}_a(x)=1/(1+4x^2)$ and $\mathcal{L}_d(x)=-2x/(1+4x^2)$ are the Lorentzian absorptive and dispersive lineshapes. The suppression of the atomic spontaneous emission results in an effective boost of the cooperativity of main cavity. In other words, a large part of the scattered photons are captured by the side cavity and interact with the atoms again. By choosing $\Gamma/\Delta \ll 1$ in the experiments, the boost factor has a simplified expression, 
\begin{align}
  \left[ N\eta_{\rm side}\frac{\Gamma}{2\Delta} \right]^2 
\end{align}
in the large $N$ limit. 
As a result, the quantum coherence time is improved to 
\begin{align}
  N g_4 T \sim \frac{1}{4}\sqrt{\eta}
  N\eta_{\rm side}\frac{\Gamma}{2\Delta}.
\end{align}
Assuming the same cooperativity of the side cavity as the main cavity 
 $\eta_{\rm side}=\eta$, this results in an $N$-independent coherence time (in the large $N$ limit)
\begin{align}
  g_4 T \sim \frac{\Gamma}{8\Delta} \eta^{3/2}.
\end{align}
{This holds in the limit of $N\eta_{\rm side} \Gamma/\Delta \gg 1$. 
Considering   $\rm ^{87}Rb$ atoms, one parameter choice is $\Gamma = 2\pi\times 6$ MHz, and $\Delta = 2\pi \times 2.4$ GHz. 
With that, we could keep $g_4 T$ to be larger than $130$ for a system of $400$ atoms, taking $\eta = 5000$, where $g_4\approx 2\pi\times 140\,\rm Hz$. 
These estimates for the cross cavity indicate the system is promising for achieving scalable quantum advantage in solving quantum optimization problems.
} 

\section{More details about atom cavity encoding for 3-SAT} 
\label{sec:3SAT} 
As introduced in the {\it main text}, 
we consider  a 3-SAT problem with $n$ variables $x_{1},\dots,x_{n}$, and $m$ clauses $f_{1},\dots,f_{m}$~\cite{Kitaev_QCBook}. 
We introduce a binary vector ${\bf x} \equiv  (x_1, \ldots, x_n)$ for a compact notation. 
The 3-SAT problem is defined by two matrices $I$ and $B$, 
with 
\be 
f_j = ( x_{I_{j,1}}\oplus B_{j,1}) 
\lor ( x_{I_{j,2}}\oplus B_{j,2} ) 
\lor  ( x_{I_{j,3}}\oplus B_{j,3}) 
\label{eq:3SATdef} 
\ee  
where $I$ contains integers from $1$ to $n$, and $B$ contains binary values. 
The 3-SAT problem is to find ${\bf x}$ that satisfies 
\be 
f_1 \land f_2 \ldots \land f_m 
\ee 

We define a set ${\cal R}$, whose number elements are 
\bea 
    && a_{i}=\sqrt{\alpha_{m+i}}+\sum_{j:x_{i} {\rm in} f_{j}}^{m}\sqrt{\alpha_{j}} \nn \\
    && b_{i}=\sqrt{\alpha_{m+i}}+\sum_{j:\overline{x_{i}} {\rm in} f_{j}}^{m}\sqrt{\alpha_{j}},   \\ 
    && c_j = d_j = \sqrt{\alpha_j} \nn
\eea 
with $\alpha_p$ the $p$-th squarefree integer, starting from $1$. We have  $i\in [1, n]$, and $j \in [1, m]$. 
The number elements of ${\cal R}$ are denoted as $r_k$ with $k \in [1, 2n+2m]$. 
Here, we shall prove that 
solving the 3-SAT problem is equivalent to  solving 
\be 
\sum_{k=1} ^{2n+2m}  y_k r_k = T
\label{eq:ykeq} 
\ee 
with $y_k$ taking binary values ($0$ or $1$), 
and $T =\sum_{i=1}^{n}\sqrt{\alpha_{m+i}}+3\sum_{j=1}^{m}\sqrt{\alpha_{j}}$. We introduce ${\bf y} \equiv  (y_1, \ldots, y_{2n+2m})$. 
Here, the equivalence means the solution of Eq.~\eqref{eq:ykeq} can be decoded from the solution of 3-SAT in Eq.~\eqref{eq:3SATdef}, and vice versa.

The proof contains two steps. First, we shall prove a forward statement---if ${\bf x} $ is a solution for the 3-SAT in Eq.~\eqref{eq:3SATdef}, then there is a corresponding solution  ${\bf y}$ (with $y_{k\in [1,n]} = x_k$) for Eq.~\eqref{eq:ykeq}. 
Second, we shall prove the reverse statement---if ${\bf y} $ is a solution for Eq.~\eqref{eq:ykeq}, then there is a corresponding solution ${\bf x}$ 
(with $x_{k\in [1,n]} = y_k$) for Eq.~\eqref{eq:3SATdef}.  
We present a constructive proof below, where the decoding recipes from ${\bf x}$ to ${\bf y}$, and from ${\bf y}$ to ${\bf x}$ are also given. 

Now, we prove the forward statement.  Suppose ${\bf x}$ is a solution for the 3-SAT in Eq.~\eqref{eq:3SATdef}, then we introduce $y_{k \in [1, n]} = x_k$, and $y_{k\in [n+1, 2n]} = \overline{x_{k-n}}$. 
We have 
\bea 
 \sum_{k=1}^{2n} y_k r_k = \sum_i x_i a_i + \overline{x_i} b_i
   =  \sum_i \sqrt{\alpha_{m+i}} + \sum_j \left\{ \sqrt{\alpha_j}  \left[ \sum_{i: x_i {\rm in} f_j} x_i +  \sum_{i: \overline{x_i}  {\rm in} f_j} \overline{ x_i}  \right] \right\}. \nn 
\eea 
Since ${\bf x}$ is a solution for all $f_j = 1$, the summation 
\[ 
l_j \equiv \left[ \sum_{i: x_i {\rm in} f_j} x_i +  \sum_{i: \overline{x_i}  {\rm in} f_j} \overline{ x_i}  \right]
\] 
takes values of $1$, $2$, or $3$ (note that each  clause in 3-SAT contains three literals). 
For the three cases with $l_j = 1$, $2$, and $3$, we choose  
($y_{2n+j} = 1$, $y_{2n+m+j} =1$), 
($y_{2n+j} =1 $, $y_{2n+m+j} = 0$), 
and 
($y_{2n+j} =0 $, $y_{2n+m+j} = 0$), respectively. For all the three cases, we have  
\[
\sum_{k=1} ^{2n+2m}  y_k r_k = \sum_{i=1}^{n}\sqrt{\alpha_{m+i}}+3\sum_{j=1}^{m}\sqrt{\alpha_{j}},
\] 
meaning Eq.~\eqref{eq:ykeq} is satisfied.

Then, we prove the reverse statement. Suppose ${\bf y}$ is a solution for Eq.~\eqref{eq:ykeq}, then we have 
\bea
&& y_{i} + y_{n+i} =1, \text{ for $i\in [1,n]$}  \nn \\ 
&& \left[ \sum_{i: x_i {\rm in} f_j} y_i +  \sum_{i: \overline{x_i}  {\rm in} f_j} y_{n+i}  \right] + y_{2n+j} +y_{2n+m+j} = 3
\eea  
for the linear independence of radicals obeyed by the squarefree integers~\cite{2007_lam_mathematics}. 
It is then guaranteed that $y_i = \overline{y_{n+i}}$. We now choose $x_i = y_i$, then we have 
\[
  \left[ \sum_{i: x_i {\rm in} f_j} x_i +  \sum_{i: \overline{x_i}  {\rm in} f_j} \overline{x_{n+i}}   \right] + y_{2n+j} +y_{2n+m+j} = 3. 
\]
It follows that   
\[ 
\left[ \sum_{i: x_i {\rm in} f_j} x_i +  \sum_{i: \overline{x_i}  {\rm in} f_j} \overline{x_{i}}   \right]  = \text{$1$, $2$, or $3$}, \text{for all $j$}. 
\] 
This implies the clauses ($f_j$) are all satisfied.

\section{Encoding for vertex cover problem.} 
Besides the 3-SAT problem and QUBO problem, we find that the  vertex cover problem, one representative graph problem, can also be encoded efficiently by the atom cavity system. 

A vertex cover of a graph is a set of vertices that touches every edge of  the graph. The vertex cover problem is to decide if a given graph $G=(V,E)$ 
{has a vertex cover of size $\kappa$.} 
The problem is NP-complete. 

We consider a graph   $G(V,E)$ with $n$ vertices [$V= (v_{1},\dots,v_{n})$], and  $m$ edges  [$E \equiv (e_{1},\dots,e_{m})$]. 
Corresponding to the vertices, we introduce $n$ numbers,  
\[
a_{i}=\sqrt{\alpha_{m+1}}+\sum_{j:i\in{e_{j}}}\sqrt{\alpha_j},i=1,\dots,n . 
\] 
Corresponding to the edges, we introduce $m$ numbers, 
\[
 b_{j}=\sqrt{\alpha_{j}}, j=1,\dots,m 
\] 
These numbers ($a_i$ and $b_j$) form a set ${\cal R}$ with size $n+m$. The number elements are denoted as $r_k$ with $k \in [1, n+m]$. 
Solving the vertex cover problem is equivalent to solving 
\be 
\sum_{k=1}^{n+m} y_k r_k  = T, 
\label{eq:ykVC} 
\ee 
with $y_k$ taking binary values ($0$ or $1$), and 
$T=\kappa\cdot\sqrt{\alpha_{m+1}}+\sum_{j=1}^{m}2\cdot\sqrt{\alpha_{j}} $. 
The proof of their equivalence is similar to the 3-SAT encoding in Sec.~\ref{sec:3SAT}. 

First, suppose the graph $G$ has a vertex cover of size $\kappa$, we prove there is a corresponding solution to Eq.~\eqref{eq:ykVC}. 
We describe the  configuration of the vertex cover by a binary vector ${\bf x}=(x_1, x_2, \ldots x_n)$, where $x_i=1$  or $0$, represents whether the $i$-th vertex belongs to the vertex cover or not. 
We then have 
\[ 
\sum_i a_i x_i = \kappa \sqrt{\alpha_{m+1}} + \sum_j \sqrt{\alpha_j} \left[ \sum_{i: i\in e_j} x_i \right]. 
\] 
By the definition of vertex cover,  the summation $l_j \equiv \sum_{i: i\in e_j} x_i$ for each edge (labeled by $j$) is $1$ or $2$. We choose ${\bf y}$ by letting $y_{i \in [1, n]} = x_i$, and 
$y_{n+j} = 1$ ($y_{n+j} = 0$) for $l_j= 1$ ($l_j = 2$). We then have 
\[
 \sum_{k=1}^{n+m} y_k r_k = \sum_i a_i x_i + \sum_j b_j y_{n+j} =  \kappa \sqrt{\alpha_{m+1}}  + \sum_j \sqrt{\alpha_j} l_j + \sum_j \sqrt{\alpha_j} y_{n+j}  \nn \\ 
= \kappa\cdot\sqrt{\alpha_{m+1}}+\sum_{j=1}^{m}2\cdot\sqrt{\alpha_{j}}. 
\]
This means Eq.~\eqref{eq:ykVC} is satisfied. 

Second, suppose Eq.~\eqref{eq:ykVC} has a solution ${\bf y}$, we prove there is a corresponding solution to the vertex cover problem. 
For  the linear independence of radicals obeyed by the squarefree integers~\cite{2007_lam_mathematics}, we have 
\bea 
\sum_{i=1}^{n} y_i  &=& \kappa \nn \\ 
\sum_{i: i\in e_j} y_i &=& \text{$1$ or $2$ for all $j$}.  
\eea  
We then select the vertices (labeled by $i$) of the graph $G$ with $y_i = 1$. This forms  a set with size $\kappa$, and 
every edge contains one or two  vertices  selected, 
which means this vortex set is a vertex cover of size $\kappa$. 

Now, we have proved that the vertex cover problem is equivalent to the problem in Eq.~\eqref{eq:ykVC}, which has the same form as the problem in  Eq.~\eqref{eq:ykeq} constructed for the 3-SAT. 
This means the vertex cover problem also maps to the atom cavity system. 
The atom number overhead for encoding the vertex cover problem is linear, 
\be 
{\rm Overhead} = m. 
\ee

\end{widetext}

\end{document}